\documentclass{aa}  
\usepackage{graphicx}
\usepackage{longtable,lscape}
\usepackage[authoryear]{natbib}
\bibpunct{(}{)}{;}{a}{}{,}
\usepackage{txfonts}
\begin{document}
%
  \title{Foregrounds for observations of the cosmological 21~cm line: I. First Westerbork measurements of Galactic emission at 150~MHz in a low latitude field} 
\author{G.~Bernardi\inst{1} \and A.G.~de~Bruyn\inst{1,2} \and M.A.~Brentjens\inst{2} \and B.~Ciardi\inst{3} \and G.~Harker\inst{1} \and V.~Jeli{\'c}\inst{1} \and L.~V.~E.~Koopmans\inst{1} \and P.~Labropoulos\inst{1} \and A.~Offringa\inst{1}\and V.N.~Pandey\inst{1} \and J.~Schaye\inst{4} \and R.M.~Thomas\inst{1} \and S.~Yatawatta\inst{1} \and S.~Zaroubi\inst{1}}
   \institute{Kapteyn Astronomical Institute, University of Groningen, PO Box 800, 9700 AV Groningen, The Netherlands\\
              \email{bernardi@astro.rug.nl}
         \and ASTRON, PO Box 2, 7990 AA Dwingeloo, The Netherlands \and Max-Planck Institute for Astrophysics, Karl-Schwarzschild-Stra\ss e, 1, 85748 Garching, Germany \and Leiden Observatory, Leiden University, PO Box 9513, 2300 RA Leiden, The Netherlands} 
   \date{Received }

 
  \abstract{The cosmological 21~cm line promises to be a formidable
  tool for cosmology, allowing the investigation of the end of the so--called dark ages,
  when the first galaxies formed.} 
  {Astrophysical foregrounds are expected to be about three
  orders of magnitude greater than the cosmological signal and
  therefore represent a serious contamination of the
  cosmological 21~cm line. Detailed knowledge of both their intensity and polarization structure on the relevant angular scale of 1--30 arcmin will be essential for extracting the cosmological signal from the data.} 
  {We present the first
  results from a series of observations conducted with the Westerbork
  telescope in the 140--160~MHz range with a 2~arcmin resolution aimed
  at characterizing the properties of the foregrounds for epoch of
  reionization experiments. The polarization data were analysed
  through the rotation measure synthesis technique. We computed total intensity and polarization angular power spectra.} 
  {For the first time we have detected
  fluctuations in the Galactic diffuse emission on scales greater than 13~arcmin at
  150~MHz, in the low Galactic latitude area known as Fan region, centred at $\alpha = 3^{\rm h}10^{\rm m}$, $\delta = 65^\circ 30'$. Those fluctuations have an $rms$ of 14~K. The total intensity power spectrum shows a power--law behaviour down to $\ell \sim 900$ with slope $\beta^I_\ell = -2.2 \pm 0.3$. The detection of diffuse emission at smaller angular scales is limited by residual point sources. We measured an $rms$ confusion noise of $\sim$3~mJy~beam$^{-1}$.
  
Diffuse polarized emission was also detected for the first time at this frequency. The polarized signal shows complex structure both spatially and along the line of sight. The polarization power spectrum is not affected by residual point sources and is only limited by the thermal noise. It shows a power--law behaviour down to $\ell \sim 2700$ with slope $\beta^P_\ell = -1.65 \pm 0.15$. The $rms$ of polarization fluctuations is 7.2~K on 4~arcmin scales.} 
  {The measured total intensity fluctuations are used to estimate the foreground contamination on the cosmological signal. By extrapolating the spectrum of total intensity emission, we find a contamination of $\delta T= \sqrt{\ell (\ell+1) C^I_\ell / 2\pi} \sim 5.7$~K on 5~arcmin scales and a corresponding $rms$ value of $\sim$18.3~K at the same angular scale.  
  
The level of the polarization power spectrum is $\delta T \sim 3.3$~K on 5~arcmin scales. However, the Fan region cannot be taken as representative of other sky regions, given its exceptionally bright polarized signal, but is likely to represent an upper limit on the sky brightness at moderate and high Galactic latitude.} 

   \keywords{Polarization -- Cosmology: diffuse radiation -- Cosmology: observations -- Radio continuum: general -- ISM: general -- ISM: magnetic fields}

   \titlerunning{WSRT observation of Galactic emission at 150~MHz}
   \maketitle
%

\section{Introduction}
Although interesting in its own right, the study of the diffuse Galactic radio emission at metre wavelengths has recently taken on more
significance because this emission is the dominant 
contaminating foreground for the study of the redshifted 21~cm line from the
epoch of reionization (EoR, Shaver et al. 1999; Furlanetto et al. 2006). The recent WMAP results on the cosmic microwave
background polarization suggest that a significant part of the 
reionization of the Universe must have occurred at z~$ < 13$, 
while the Gunn-Peterson effect in high--redshift quasars
tells that the bulk of the reionization  must have occurred earlier than
z$\sim$6. The best frequencies for searching for the cosmological
21~cm line therefore lie in the 100--200~MHz range. 

The 21~cm line is expected to have an $rms$ brightness
temperature of a few millikelvin on angular scales of a few arcmin
(Zaldarriaga et al. 2004). However, both the angular 
scale and the intensity depend on the details of the reionization
process. Simulations indicate that the cosmological signal
is expected to vary with redshift (Ciardi \& Madau 2003; Mellema et al. 2006; Thomas et al. 2008).

The Giant Metrewave Radio Telescope (GMRT) and several dedicated experiments like LOFAR\footnote{http://www.lofar.org}, 
MWA\footnote{http://haystack.mit.edu/ast/arrays/mwa} and
21CMA\footnote{http://web.phys.cmu.edu/past} will search for these
signals. 

The detection of cosmological signals with millikelvin brightness temperatures
is plagued by a wide range of challenges. Foremost among these is the
intense foreground emission from our Galaxy (Shaver et al. 1999).  In fact about 75\% of the sky brightness is due to
diffuse emission from our Galaxy, most of which is caused by the
synchrotron process. Our present knowledge of the Galactic diffuse
radio emission at metre wavelength is not adequate to estimate the
contaminating effects on the EoR signal. Several all-sky maps (Haslam
et al. 1982; Reich \& Reich 1986; Reich et al. 2001; Page
et al. 2007) at different frequencies and angular resolutions are
available, but the radio emission from our Galaxy at the relevant
frequencies (100-200~MHz) is poorly known. The 150~MHz single--dish all-sky map by
Landecker \& Wielebinski (1970) has only $2^\circ$ resolution, while estimates of the spatial fluctuations on arcmin scales are required in EoR experiments.

The first indications that the Galactic
foreground lacks significant fine-scale structure at low frequencies
were presented by Wieringa et al. (1993). Subsequent dedicated
arcmin resolution observations with the Westerbork telescope (WSRT) at frequencies from
315 to 380~MHz, have indeed shown that the diffuse synchrotron emission
from the Galaxy is generally very smooth and therefore mostly resolved out
by the interferometer. This also became evident from the Westerbork Northern Sky Survey (WENSS) at
325~MHz (Rengelink et al. 1997, see also WoW\footnote{http:/www.astron.nl/$\sim$wow}). 

Extrapolating these limits to lower frequencies and smaller angular
scales is uncertain because both the spectral and the spatial
properties of the diffuse emission can vary across the sky. The mean
spectral index of the synchrotron emission at high Galactic latitude
has been quite well constrained to be $\beta
= 2.5 \pm 0.1$ in the 100--200~MHz range (Rogers \& Bowman 2008). 

The spatial properties are less well known because no power spectra of the diffuse emission exist on arcmin scales. 
La Porta et al. (2008) analysed power spectra of the diffuse 
Galactic synchrotron radiation down to scales of a few degrees and 
found significant variations across the sky even after subtracting 
the brightest sources.

Although the redshifted 21~cm line signal is itself unpolarized, 
we still have to worry about intrinsically polarized
signals from the sky. The instrumental polarization of low frequency
radio telescopes and the upcoming arrays is significant (10--20\%). 
Imperfectly calibrated polarized signals from the sky will then leak 
into the total intensity images. Even small variations in the
rotation measure of the polarized signals will then lead to frequency
dependent residuals, corrupting the feeble cosmological 21~cm signal. It is
therefore prudent to search for EoR target fields with low
foreground polarization. 

Until recently, most studies of Galactic polarization were done at
relatively high frequencies. Brouw \& Spoelstra (1976) collected data with the Dwingeloo telescope, which cover large
regions of the sky in the 408--1411~MHz range with a resolution between
0.5$^\circ$ and $2^\circ$. A full--sky absolutely--calibrated polarization map at
1.4~GHz with 0.5$^\circ$ resolution was recently published by Wolleben
et al. (2006).  Selected regions of low total intensity synchrotron
emission have been observed between 1.4 and 2.3~GHz (Bernardi et al. 2003; Carretti et
al. 2005a; Carretti et al. 2005b; Carretti et al. 2006; Bernardi et
al. 2006) characterizing the spatial properties of the diffuse
polarized emission down to arcmin scales with the purpose of studying
the contamination of the cosmic microwave background polarization.

At lower frequencies, the sky coverage has been scarce.  Following 
the pioneering WSRT work by Wieringa et al. (1993) at 325 MHz, several dedicated 350~MHz 
observations in selected regions have been conducted 
(Haverkorn et al. 2003a; Haverkorn et al. 2003b - hereafter HKB) together with observations of a limited number of 
WENSS survey fields (Schnitzeler 2008). These studies indicated that foreground
polarization levels, even at higher Galactic latitudes, could easily go up to 5~K or more on 5--10~arcmin scales (de Bruyn et al. 2006). If these regions were
Faraday thin, the polarized brightness temperatures at 150~MHz could go up 
to many tens of kelvin. This would place extraordinary requirements on the
level of accuracy of the instrumental polarization calibration.  

Several authors have simulated the properties of the foregrounds in order
to test strategies and algorithms to subtract or filter them from the
data (Di Matteo et al. 2004; Morales et al. 2006; Gleser et al. 2007;
Jeli{\'c} et al. 2008, Liu et al. 2009a, Harker et al. 2009, Liu et al. 2009b). It is clear, however, that these simulations require
guidance from observational data on both total and polarized foreground levels.

We have therefore initiated a comprehensive programme to directly measure the
properties of the Galactic foreground in the 100--200~MHz range, the relevant frequencies for the EoR. We used the Low Frequency Front Ends (henceforth LFFE) on the WSRT to carry out these observations.  We chose three
different fields. One field is centred around the so--called Fan
region, at Galactic coordinates $l=137^\circ$ and $b=+8^\circ$ in the 2nd Galactic quadrant. It
was selected because it remains highly polarized down to the 350~MHz frequencies
(Brouw \& Spoelstra 1976; Haverkorn et al. 2003a).  The other two fields 
were chosen because they represent possible targets for EoR
observations. The second field is centred on the very
bright radio quasar 3C196 in one of the coldest regions of the
Galactic halo. The third field is located close to the North
Celestial Pole, which would allow us to collect night time observations 
throughout the year at the geographic latitude of the LOFAR array
(+53$^\circ$).

In this paper we present results from WSRT observations of
the Fan region at 150~MHz aimed at studying the properties of both the total
intensity and the polarized diffuse emission as contaminants for the
EoR signal. In subsequent papers we will present results from the
analysis of the other fields, together with the challenges and the
calibration strategies for foreground subtraction applied to the WSRT
data.

This paper is organized as follows: in Section~\ref{obs_res} we present
the observations together with the data reduction and the basic
results, in Section~\ref{diff_gal} we describe the results of the
total intensity emission, in Section~\ref{pol_em} we analyse the
polarized emission, in Section~\ref{power_spec} we present the power spectrum analysis and in Section~\ref{concl} we conclude.

\section{Observations and data reduction}
\label{obs_res}

Observations of the Fan region took place between the end of November and the beginning
of December 2007 for a total of $6 \times 12$~hours. All observations of the
target field took place between sunset and dawn, avoiding corruption of
the shortest baselines by the strong and often time--variable emission 
from the Sun. Short calibration observations of the highly polarized
pulsar J0218+4232 (Navarro et al. 1995) and the strong unpolarized
source 3C196 were made immediately before and after the target
observation, which proceeded uninterrupted. 

The WSRT telescope consists of 14 dishes of 25~m diameter, ten of
which (labelled  0 to 9) are on fixed locations 144~m apart. The other four (labelled  A to D) are movable on a rail track. The redundant baselines are normally not included in the imaging process, because they cause very strong grating lobes. The normal choice is to use all the correlations between the ten fixed antennas and the four movable ones. 
Using six configurations of the movable telescopes, the four movable dishes were incrementally moved by 12~m obtaining a uniform $uv$ coverage from 36~m up to the maximum spacing of 2760~m.
Unfortunately antenna 5 was missing in all the observations,
causing two broad gaps in the $uv$ plane centred at $\sim 300$~$|{\bf u}|$ and
$\sim 900$~$|{\bf u}|$, where $|{\bf u}|=\sqrt{u^2+v^2}$ is the projected distance in the $uv$ plane measured in wavelengths.
\begin{table}	
\caption[]{Summary of the observational setup.} 
\label{obs_table}
\begin{tabular}{l l}        
\hline\hline 
   
Coordinates of the field 		& \\
centre (J2000.0) 			& $\alpha = 3^{\rm h}10^{\rm m}$, $\delta = 65^\circ 30'$\\
Number of spectral bands	  	& 8\\
Central frequency of each band (MHz)	& 139.3, 141.5, 143.7, 145.9,\\ 
					& 148.1, 150.3, 152.5, 154.7\\
Width of each band (MHz)		& 2.5\\
Frequency resolution (kHz)		& 4.9 (9.8 after tapering)\\
Time resolution (sec)			& 10\\
Angular resolution			& $2'\times 2' \rm{cosec}(\delta) \simeq 2' \times 2.2'$\\
Conversion factor			& 1~mJy~beam$^{-1}$ = 3.98~K\\
\hline
\end{tabular}
   \end{table}

Table~\ref{obs_table} summarizes the main characteristics of the
observations. The eight spectral windows were chosen to provide a
contiguous frequency coverage, allowing for a small overlap between
the bands. A large part of the fourth spectral band, centred at 
145.9~MHz, turned out to be too severely contaminated by radio
interference and the whole band was discarded from the analysis. 

After Hanning tapering the 512 channels, and discarding the edges of each
band, we were left with a total of about 240 frequency channels per band.
Each channel has a width of 9.8~kHz. 

The data were reduced using the
AIPS++\footnote{http://aips2.nrao.edu/docs/aips++.html} package. We
integrated the standard AIPS++ distribution with flagging routines
explicitly developed for dealing with radio frequency interference (RFI) and with routines which determine the polarization
calibration. Dipole gain and leakage corrections were
determined using the unpolarized calibrator 3C196. Due to an 
unknown phase offset between the horizontal and the vertical dipoles,
signal can leak from Stokes $U$ into Stokes $V$ (Sault, Hamaker \& Bregman 1996). In order to correct for this phase difference, the
polarized pulsar PSRJ0218+4232 was observed for about 15~min just prior to the
start of the 12~hour synthesis. Since the pulsar has a known rotation measure $RM = -61$~rad~m$^{-2}$ (Navarro et al. 1995), we corrected the phase difference by rotating the polarization vector in the plane defined by the Stokes $U-V$ parameters in order to have zero Stokes $V$ flux, where the direction of the rotation has to provide a negative RM for the pulsar.

The flux scale at low frequencies is not very accurately determined and a more definitive flux scale determination is being developed for LOFAR. In our case we used the radio source 3C196 as the primary WSRT flux calibrator at low frequencies. 3C196 is a very bright steep spectrum 
radio galaxy whose spectrum appears smooth down to about 10 MHz (Laing \& Peacock 1980). We have adopted a flux density at 150 MHz of 76.8~Jy for 3C196 and a power-law spectral index of $\alpha=-0.64$ in the frequency range from 115--175~MHz -- although there may be a slight steepening at the upper end of this band. We believe this number to be accurate to $\sim$5\%. 

The system noise is constituted by the receiver noise and the brightness temperature of the sky itself. At low frequency our Galaxy contributes significantly to the system noise. 3C196 is located in the coldest spot of the Galactic halo according to the 408~MHz all--sky map (Haslam et al. 1982). The System Equivalent Flux Density (SEFD) of the telescopes at 150~MHz is about 7000~Jy in the 3C196 area but it rises to over 10,000~Jy in the Galactic plane where the Fan region is located. Because the WSRT receivers
operate with an automatic gain control system before the analog--to--digital converter, it continuously measures the total power to allow corrections for the variable input levels. 

Since the total power detectors (which integrate the power over the whole 2.5~MHz band) are corrupted by RFI for most of the time, we cannot automatically correct the correlation coefficients for the variations in the system noise. The LFFE band is full of mostly impulsive and narrow band RFI coming from airplanes, satellites and mobile users as well as electronic hardware within the building which is located halfway along the array. At the high spectral resolution of 9.8~kHz provided by the backend most of this RFI can be excised. However, the total power data must be manually inspected for suitable stretches of power level measurements. These data form the basis for a manual correction of the flux scale.

We found that the total power ratio between the FAN region and the 3C196 field is $1.48 \pm 0.03$ over the 138--157~MHz band. Following the transfer of the bandpass correction determined for 3C196 we have therefore applied an additional correction of a factor 1.5 to the visibility data. 

After this correction was applied, the data were self--calibrated in order to correct for time variations in the dipole gains. Self-calibration is especially useful at metre wavelength because radio interference causes temporal variations in the gains and ionospheric turbulence introduces temporal variations in the visibility phases.

The signal to noise ratio (SNR) in our data is quite poor, given that the brightest source is $\sim$2.8~Jy and the noise per visibility, per channel, per time slot and per polarization is $\sim$30~Jy. Moreover we want to avoid averaging the visibilities in time to improve the SNR because a time average decorrelates the visibilitity signal if the ionosphere is turbulent on short time scales. We therefore adopted the following procedure to calibrate the data.

After the bandpass correction determined from 3C196 was applied, we generated frequency cube images for every night and every spectral band. We normalized the dirty images to the average peak value derived from every cube. In this way we compensated for frequency dependent amplitude errors.

Cyg~A and Cas~A were then subtracted from the data because at these frequencies they are bright enough to generate significant side lobes within the field of view -- at a level of $\sim$150~mJy for Cas~A -- despite being far outside the target field. Afterwards we created a sky model per night and per spectral window which contained the brightest 70-80 sources, down to a flux limit of $\sim$300~mJy. For every night and for every spectral window we then averaged all the channels to improve the SNR, and computed phase solutions every two minutes to correct for ionospheric turbulence. After this phase correction was applied, the data in the spectral bands centred at 143.7 and 148.1~MHz still showed amplitude variations as a function of time. We corrected for amplitude variations in the complex gains by computing solutions every five minutes. 
 
\begin{figure*}
   \centering
   \resizebox{1.0\hsize}{!}{\includegraphics{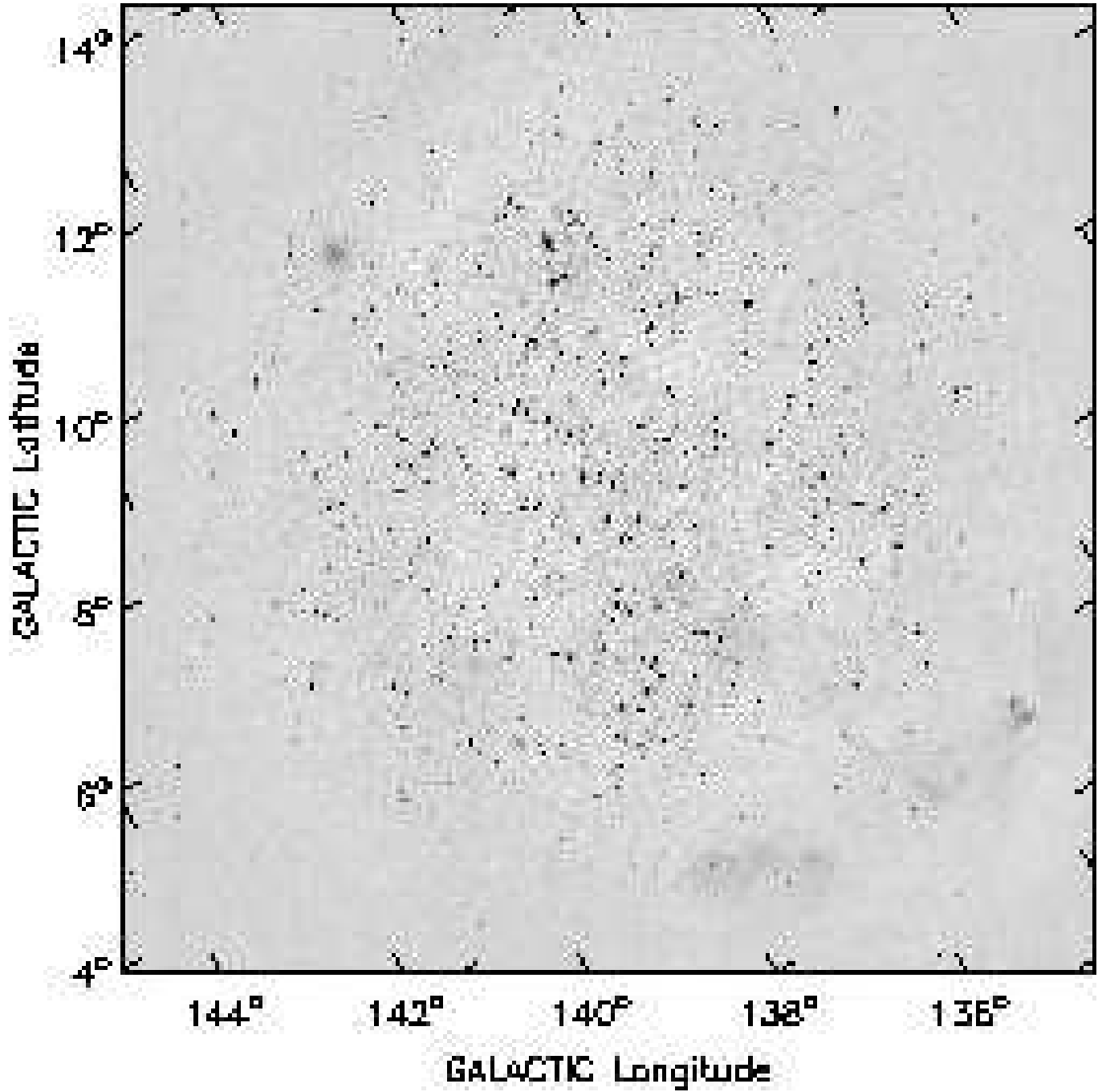}}
   \caption{Stokes $I$ map obtained by averaging all the data. The colour scale saturates (black) at 75~mJy. The tick marks indicate lines of constant Galactic latitude and longitude}
              \label{fan_stokes_spw1}
    \end{figure*}

The resulting Stokes $I$ map obtained by averaging all the seven spectral windows is shown in Figure~\ref{fan_stokes_spw1}. Since the Half Power Beam Width (HPBW) of
the WSRT dish is $\sim$6$^\circ$ at 150~MHz, Figure~\ref{fan_stokes_spw1}
maps the full primary beam. Unless stated otherwise, none of the the images shown in this paper are corrected for the primary beam attenuation. The noise is therefore constant across the images, but the celestial source brightness is modulated by the radial dependence of the beam gain.

The image is mostly dominated by emission
from unresolved point sources. Faint diffuse emission is 
visible in the South-West corner of the image where the Galactic Plane
is located. The brightest emission regions, which are heavily attenuated by the
primary beam, are due to HII regions and diffuse structures within  the Perseus arm located near Galactic coordinates $b \sim 1^\circ$ and $133^\circ < l < 138^\circ$.

%
%

\section{Diffuse Galactic foreground}
\label{diff_gal}

\begin{figure*}
   \centering
   \resizebox{1.0\hsize}{!}{\includegraphics[angle=-90]{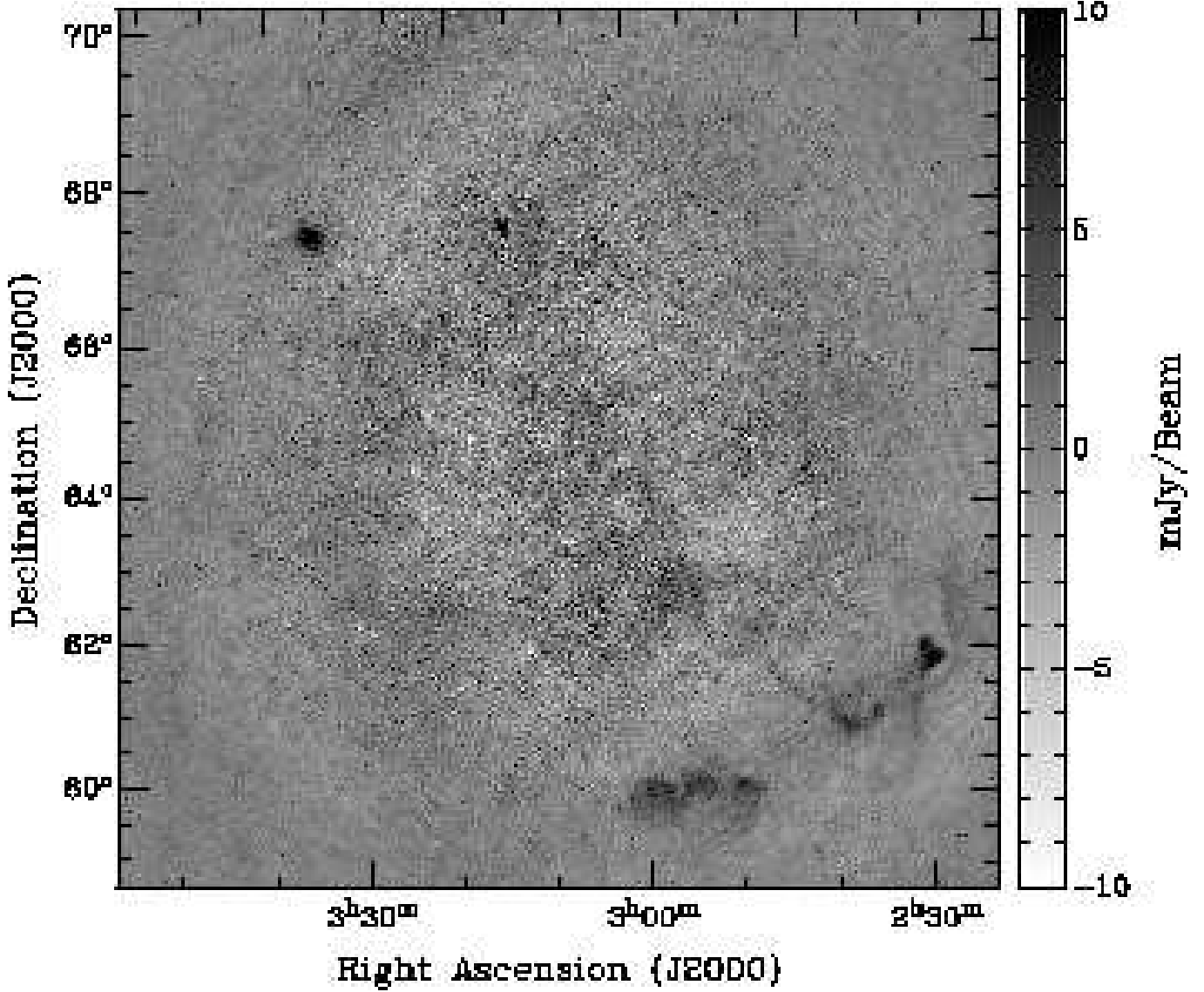}}
   \caption{Stokes $I$ map where the point sources were removed down to a level of 15~mJy level. The conversion factor is 1~mJy~beam$^{-1}$ = 3.98~K.}
              \label{full_stokes}
    \end{figure*}

Since our main interest is to characterize the fluctuations in the diffuse
Galactic foreground, we removed the point sources from the
map. Several thousand sources were identified through a CLEAN
deconvolution in a $12^\circ \times 12^\circ$ image. The CLEAN model, containing all the sources brighter
than 15~mJy, was then subtracted from the visibility
data. Figure~\ref{full_stokes} shows the residual image made from the
sum of seven bands. The diffuse structure in the Perseus
arm now becomes clearly visible. The nearby galaxy IC342 is also
noticeable as a bright structure around $\alpha =3^{\rm h} 38^{\rm m}$ and $\delta=+67^\circ 30'$.

Figure~\ref{full_stokes} still shows error patterns such as rings of negative values around the brightest sources, indicating that the calibration is not yet perfect on them. Moreover, spiky error patterns are also present around bright sources, which may be due to non-isoplanaticity errors caused by ionospheric phase fluctuations. A direction-dependent calibration will be applied in the future to try to correct for these errors. 

We can measure the noise from the image itself by taking the $rms$ value at the edge of the field, where the primary beam attenuation has reduced the sky brightness substantially. The observed noise at the edge of the image is 0.75~mJy~beam$^{-1}$, making these observations the deepest available to date at these frequencies.

In the map of Figure~\ref{full_stokes}, diffuse structure on scales of 15--30~arcmin is clearly visible. However, it is not clear whether diffuse structure exists also on arcmin scales. In order to investigate this point, we made an image which only contains the high spatial frequencies by selecting the baselines with $|{\bf u}| > 150$~wavelengths for each spectral band.

\begin{figure*}
   \centering
   \resizebox{1.0\hsize}{!}{\includegraphics[angle=-90]{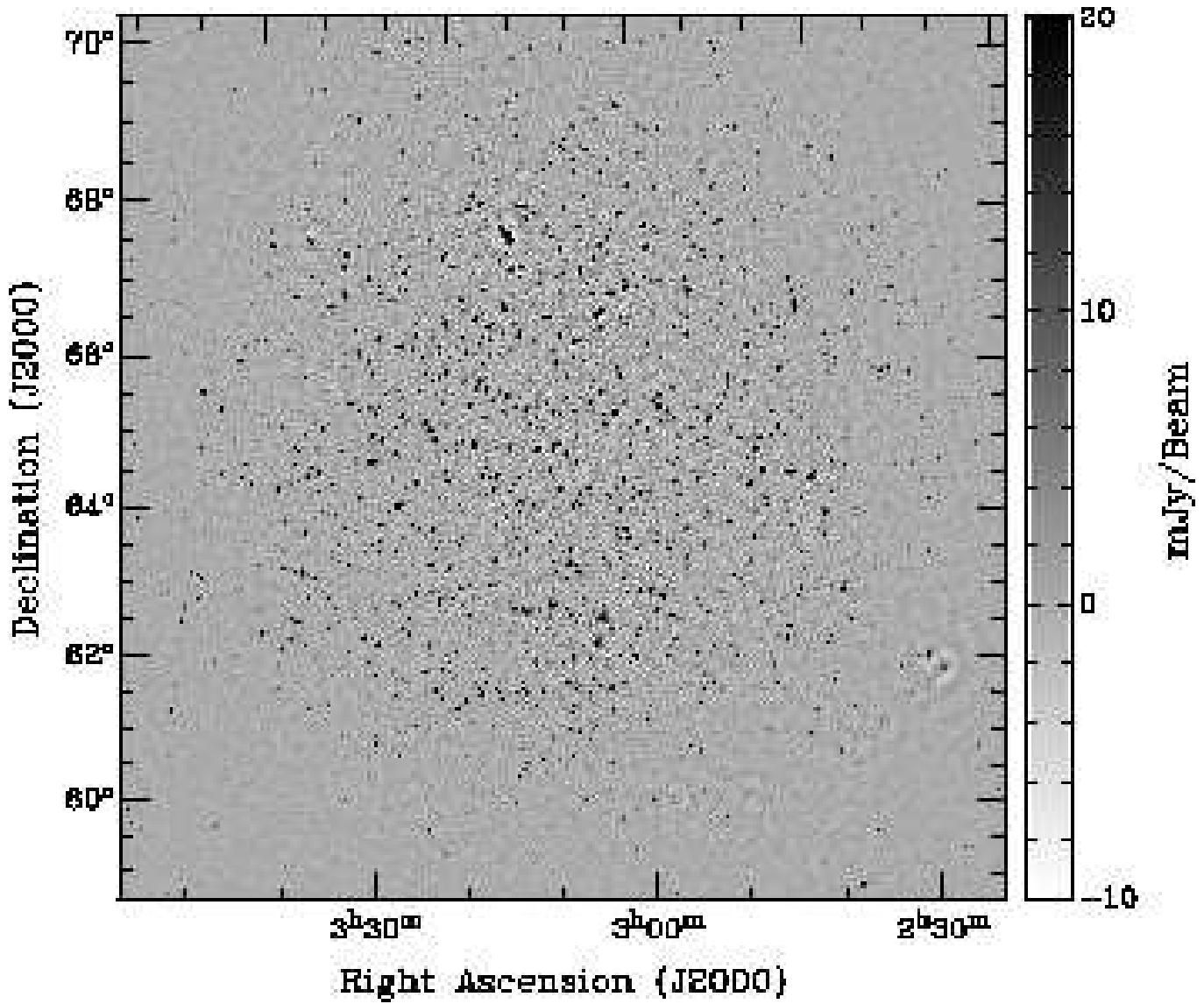}}
   \caption{Stokes $I$ map made by including only the baselines with $|{\bf u}| > 150$~wavelengths. The map has been restored after a CLEAN deconvolution. The conversion factor is 1~mJy~beam$^{-1}$ = 3.98~K.}
              \label{fan_high_res}
    \end{figure*}

The result is shown in Figure~\ref{fan_high_res} where all the angular scales greater than $\sim$12~arcmin are absent: the diffuse structure appears almost completely filtered out and the most relevant contribution left is from point sources.
This indicates that the diffuse structure appearing in the map of Figure~\ref{full_stokes} is on scales greater than 10--12~arcmin and no diffuse power is present on smaller scales.

We estimated the $rms$ due to residual point sources by subtracting from the high resolution image of Figure~\ref{fan_high_res} all the sources down to 15~mJy. We considered the inner $6^\circ \times 6^\circ$ area corrected for the WSRT primary beam $A(f,\gamma)$ which is approximated as:
\begin{eqnarray}
    A(f,\gamma) = \cos^6(0.065 \, f \, \gamma)
\label{primary_beam}
\end{eqnarray}
where $f$ is the observing frequency in MHz and $\gamma$ is the angular distance from the pointing centre in radians. 

\begin{figure}
   \centering
   \resizebox{\columnwidth}{!}{\includegraphics[angle=-90]{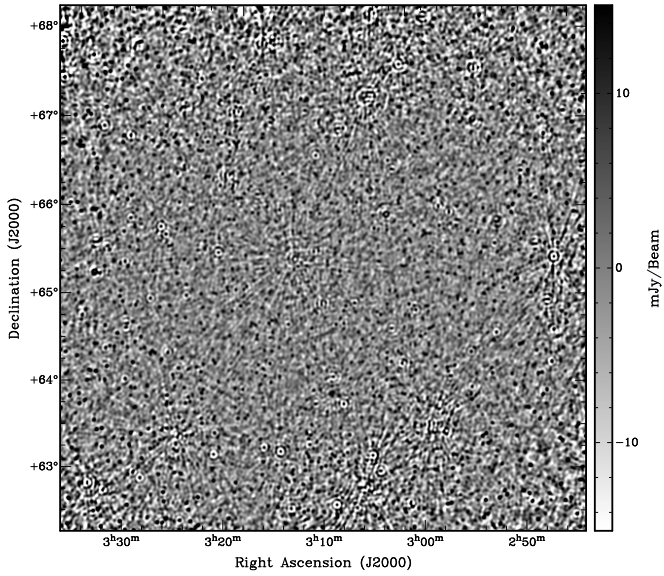}}
   \caption{$6^\circ \times 6^\circ$ Stokes $I$ map made by including only the baselines with $|{\bf u}| > 150$~wavelengths and where the point sources were subtracted down to a level of 15~mJy. The conversion factor is 1~mJy~beam$^{-1}$ = 3.98~K. The image has been divided by the primary beam according to Eq.\ref{primary_beam}.}
              \label{fan_point_sources}
    \end{figure}

Figure~\ref{fan_point_sources} shows the resulting image where the point sources have become indistinguishable from the background. This map has an $rms$ value $\sigma_{\rm{ps}} = 7.2 \pm 0.3$~mJy~beam$^{-1}$. By taking the inner $3^\circ \times 3^\circ$ area where the primary beam correction is still small, we determined an $rms$ value of $\sim$4.5~mJy~beam$^{-1}$. However, if we carefully select regions which appear to be free of instrumental or ionospheric residuals we find an $rms$ of $\sim$3~mJy~beam$^{-1}$.

This number is consistent with the expected confusion noise that can be estimated from source counts at higher frequencies. Deep WSRT observations at 1.4~GHz of the Hubble Deep Field measured a confusion noise of 5~$\mu$Jy~beam$^{-1}$ (Garrett et al. 2000). By assuming an average spectral slope of $\alpha = -0.8$ for the sources, the confusion noise $\sigma^{150}_{\rm conf}$ expected at 150~MHz can be predicted from the confusion noise at 1.4~GHz as: 
\begin{eqnarray}
\sigma^{150}_{\rm conf} = \sigma^{1400}_{\rm conf} \left( \frac{150}{1400} \right)^{-2.8} \sim 3 \, \rm {mJy~beam}^{-1}.
\label{conf_noise}
\end{eqnarray}
We can therefore conclude that the map of Figure~\ref{full_stokes} is confusion limited towards its centre rather than limited by thermal noise.

\begin{figure*}
   \centering
   \resizebox{1.0\hsize}{!}{\includegraphics{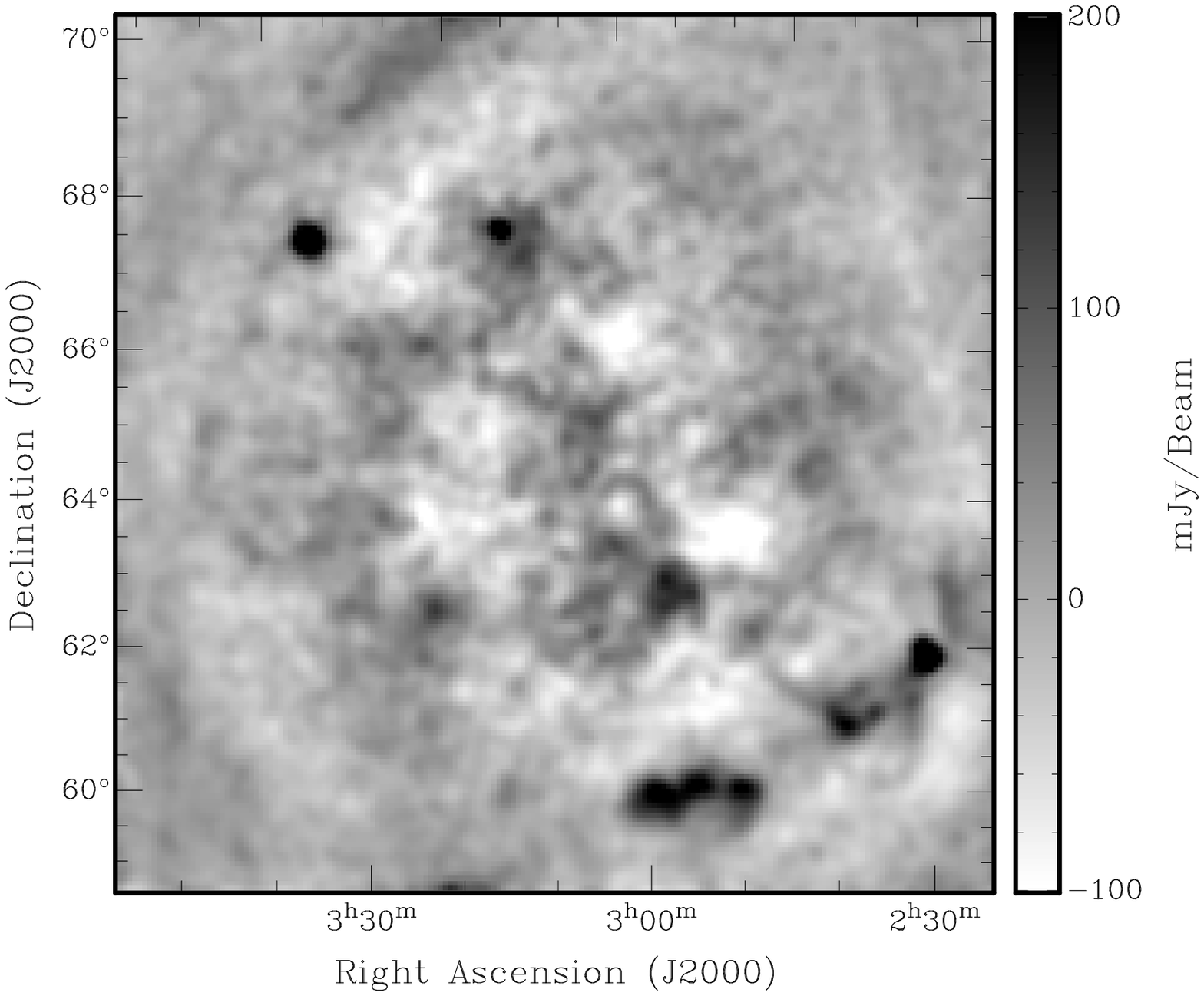}}
   \caption{Stokes $I$ map where the point sources were removed down to a level of 15~mJy and the $uv$ plane was tapered to achieve an angular resolution of 13~arcmin. The conversion factor is 1~Jy~beam$^{-1}$ = 105.6~K.}
              \label{large_scale}
    \end{figure*}

Afterwards, we characterized the diffuse emission by applying a $uv$ taper to degrade the resolution to $\sim$13~arcmin for each of the 7~bands after all the sources down to 15~mJy were removed. The resulting image is shown in Figure~\ref{large_scale}.
%
%
%
%
%
%
%
%

In the 13~arcmin image the contribution from individual point sources has completely disappeared whereas the pattern of fluctuations dominates the whole image. This pattern does not correlate with the distribution of point sources shown in Figure~\ref{fan_high_res}, indicating that possible unsubtracted sources do not contribute significantly to the diffuse emission. 

Considering that the thermal noise is reduced to less than 1~K and that the conversion factor is now 1~Jy~beam$^{-1}$~$\sim$105.6~K, the hot spots in the image are more than ten times above the noise. We conclude that the observed fluctuations represent structure in the Galactic foreground radio emission.

We characterize the fluctuations through their $rms$ value computed in the inner $6^\circ \times 6^\circ$ region of the map, after correcting for the power pattern of the primary beam and avoiding the local bright features which are definitely not representative of the diffuse foreground emission. In Table~\ref{rms_values} we report the $rms$ values of the residual maps at full and low resolution. Errors on the $rms$ values are computed by splitting the field into two halves and taking the average of the differences between these two values and the $rms$ of the whole field.

It is important to note that the $rms$ value on 13~arcmin accounts for diffuse emission only whereas the value at 2~arcmin has a significant contribution from unsubtracted point sources. 
\begin{table}	
\caption[]{$Rms$ fluctuations in the Fan field as a function of the angular resolution.} 
\label{rms_values}
\centering
\begin{tabular}{c c c}        
\hline\hline 
   
Angular resolution (arcmin) 	& $rms$ value (K)	& $rms$ value (mJy~beam$^{-1}$)\\
\hline
2				& $30.5 \pm 0.9$	& $7.7 \pm 0.3$\\
13				& $14 \pm 1$		& $128 \pm 9$\\

\hline
\end{tabular}
\end{table}

\begin{figure}
   \centering
   \resizebox{1.0\hsize}{!}{\includegraphics{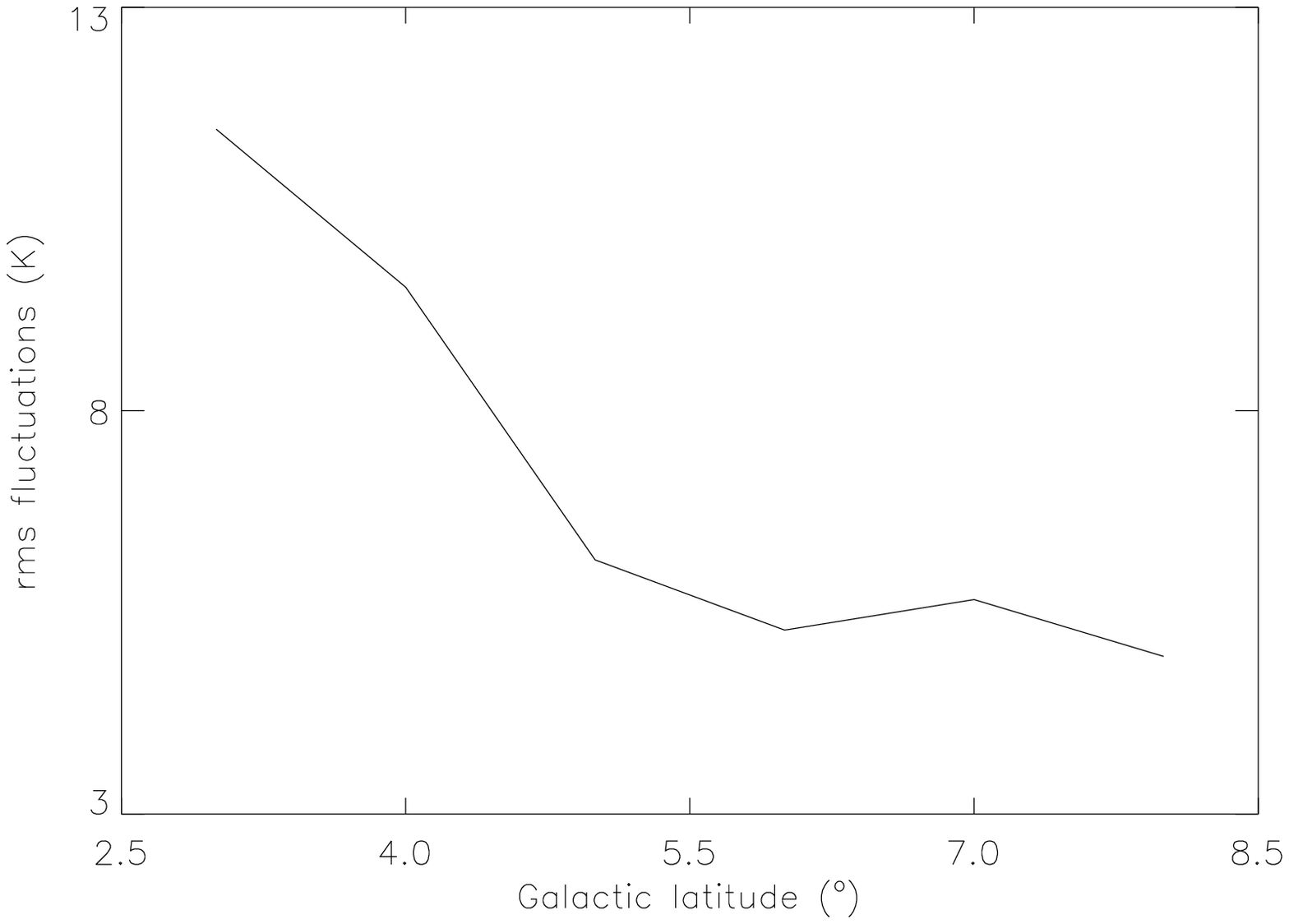}}
   \caption{$Rms$ fluctuations at 13~arcmin as a function of Galactic latitude, measured in bins of width $1^\circ$ of Galactic latitude.}
              \label{gal_plot}
    \end{figure}

Finally, we analysed the behaviour of the diffuse emission as a function of the Galactic latitude. In all Galactic radio surveys, the amplitude of the fluctuations
decreases at high Galactic latitudes. Since our data span several
degrees just outside the Galactic plane, we computed the $rms$ as a function of Galactic latitude. The result is
shown in Figure~\ref{gal_plot}. 

\begin{figure}
   \centering
   \resizebox{1.0\hsize}{!}{\includegraphics{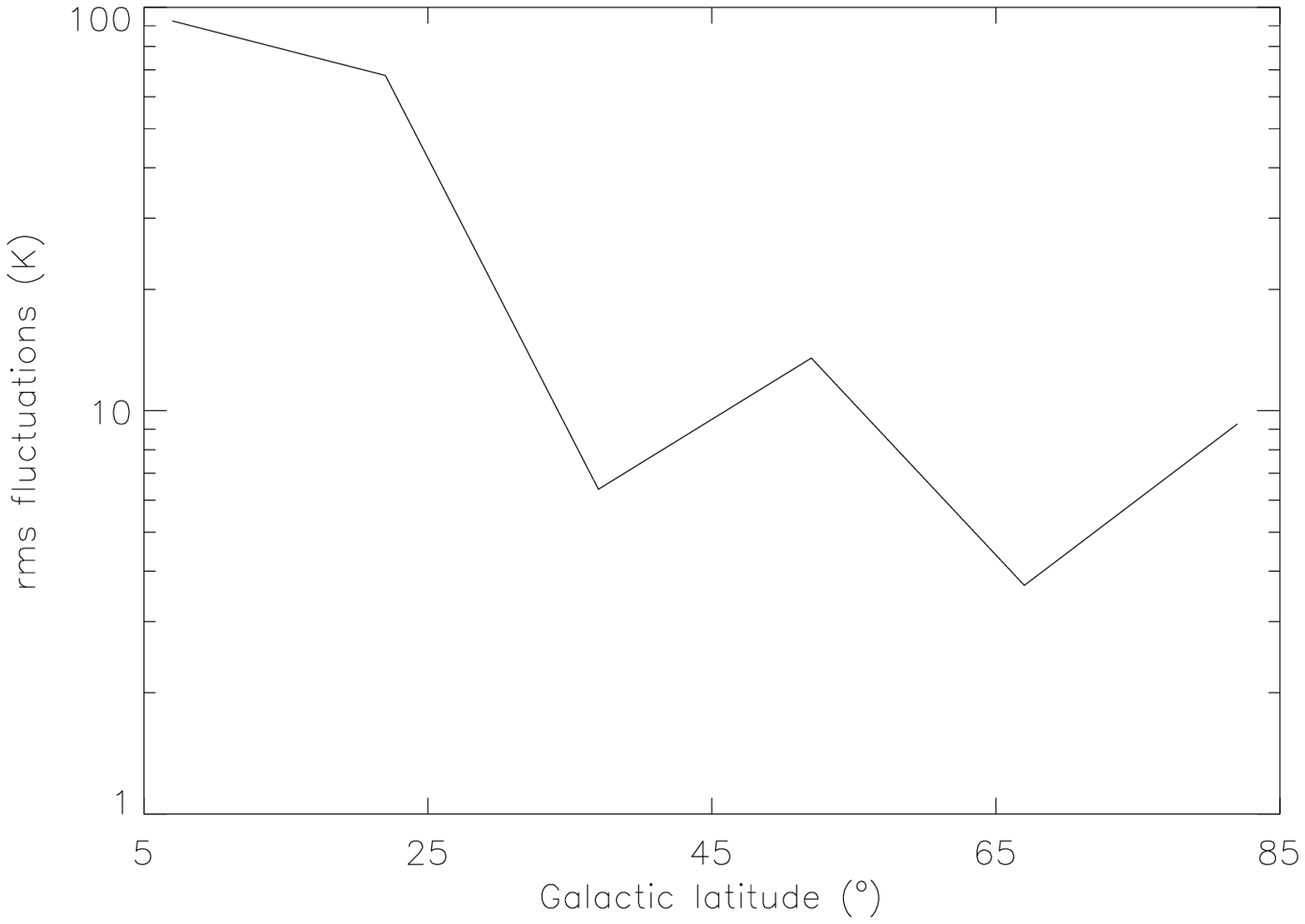}}
   \caption{$Rms$ fluctuations as a function of the Galactic latitude measured from the map by Landecker \& Wielebinski (1970). The data were grouped into bins of size $15^\circ$ of Galactic latitude and $12^\circ$ of Galactic longitude.}
              \label{gal_land_plot}
    \end{figure}

There is a clear drop in the signal as we move away from the plane. The $rms$ decreases by a factor of $\sim$2.3 over $6^\circ$ of Galactic latitude. In Figure~\ref{gal_land_plot} we plot the same behaviour derived from the Landecker \& Wielebinski map at 150~MHz which has a $2^\circ$ resolution. We considered a strip $12^\circ$ wide in Galactic longitude, centred on the Fan region, and we computed the $rms$ as a function of Galactic latitude. The $rms$ drops by a factor of ten between low and moderate Galactic latitudes ($b > 30^\circ$), staying somewhat constant at higher Galactic latitude. We expect that the sub-degree fluctuations that we have observed follow a similar decrease at moderate and high Galactic latitudes.

It is therefore relevant to notice that the same region of sky in the map by Landecker \& Wielebinski shows $rms$ fluctuations of $\sim 90$~K on $2^\circ$ scales. Since their data come from single dish observations, they sample all the angular scales down to the beam resolution. The fact that our interferometric data filters out most of the large scale power causes the $rms$ to drop by a factor of $\sim$6.5 from $2^\circ$ to 13~arcmin scales.

%
%
%
%

\section{Polarized emission}
\label{pol_em}

As we mentioned in the introduction, the Fan region is known to be
polarized from its first observations made with the Dwingeloo
telescope. Those data already showed a certain uniformity in the
polarization angles in this area. HKB observed the Fan
region at 350~MHz and found significant polarized emission on scales
of 5~arcmin.

We analysed the polarization data at 2~m wavelength through the
rotation measure synthesis technique (Brentjens \& de Bruyn 2005). The RM synthesis technique takes advantage of the Fourier relationship which exists
between $P(\lambda^2)$ and $F(\phi)$:
\begin{eqnarray}
    P(\lambda^2) = W(\lambda^2) \int^{+\infty}_{-\infty} F(\phi) \, \rm{e}^{2i\phi\lambda^2} \rm{d} \phi
\end{eqnarray}
where $P(\lambda^2)$ is the complex polarized surface brightness, $W(\lambda^2)$ is a weighting function and $F(\phi)$ is the complex polarized surface brightness per unit of Faraday depth, $\lambda$ the observing
wavelength and $\phi$ is the Faraday depth (see Brentjens \& de Bruyn 2005 for a full description).
The RM synthesis performs a transformation from the $\lambda^2$ space to the Faraday depth space. The output of the RM synthesis analysis is a cube of polarized maps at selected values of Faraday depth. The Fourier transform of  $W(\lambda^2)$ gives the RM spread function (RMSF), which is the resolution in Faraday depth and depends upon the difference in $\lambda^2$ between the two furthermost channels. The RMSF is the one--dimensional analogue of the Point Spread Function in traditional interferometry.

Table~\ref{RM_table} summarizes the most relevant parameters for the polarization analysis. The first three parameters reported in  Table~\ref{RM_table} are defined by the relationships (61), (62) and (63), of Brentjens \& de Bruyn (2005): $\delta \phi$ is the resolution in Faraday depth, $\phi^{\rm scale}_{\rm max}$ is the maximum scale of Faraday depth to which our observations are sensitive and $||\phi_{\rm max}||$ is the maximum RM value measurable.

It is worth noting that our observations are only sensitive to Faraday depths which are smaller than the RMSF width, so that we lack sensitivity to structures which are extended in Faraday depth and we are able to sample only Faraday thin regions, with an extension in RM space of $\sim$1~rad~m$^{-2}$. 
\begin{table}	
\caption[]{Summary of RM synthesis cube parameters} 
\label{RM_table}
\centering
\begin{tabular}{l l}        
\hline\hline \noalign{\smallskip}
   
$\delta \phi$				& 3~rad~m$^{-2}$\\
$\phi^{\rm scale}_{\rm max}$		& 0.85~rad~m$^{-2}$\\
$||\phi_{\rm max}||$			& 2650~rad~m$^{-2}$\\ 
Angular resolution			& 4.2~arcmin\\
Conversion factor			& 1~mJy~beam$^{-1}$ = 1~K\\
\hline
\end{tabular}
\end{table}

Although we corrected for the on-axis instrumental polarization, the WSRT has a strong off-axis instrumental polarized response which increases with distance from the centre of the image and can reach up to 20--30\% at the beam half--power radius. The off-axis polarization is also strongly frequency dependent and the polarization beam pattern is quite complicated at these frequencies. Our sky model for polarization analysis is therefore somewhat different from the total intensity model. 

We identified all the point sources down to 150~mJy in each single channel image and fitted their flux under the assumption that they were instrumentally polarized, i.e. they did not have intrinsic polarization. These sources were then subtracted from the data. In addition, Cyg~A and Cas~A were modelled and subtracted per frequency channel to properly deal with the frequency dependence of the instrumental polarization.

Since most of the Galactic polarized emission appears on spatial scales of a few arcmin, we make low resolution polarized images by discarding
the antennas 0, C and D. With this selection the array has $\sim$4.2~arcmin resolution, 
where the conversion factor at this resolution is 1~mJy~beam$^{-1}$ $=$ 1~K.

\begin{figure*}
   \centering
   \resizebox{0.9\columnwidth}{!}{\includegraphics{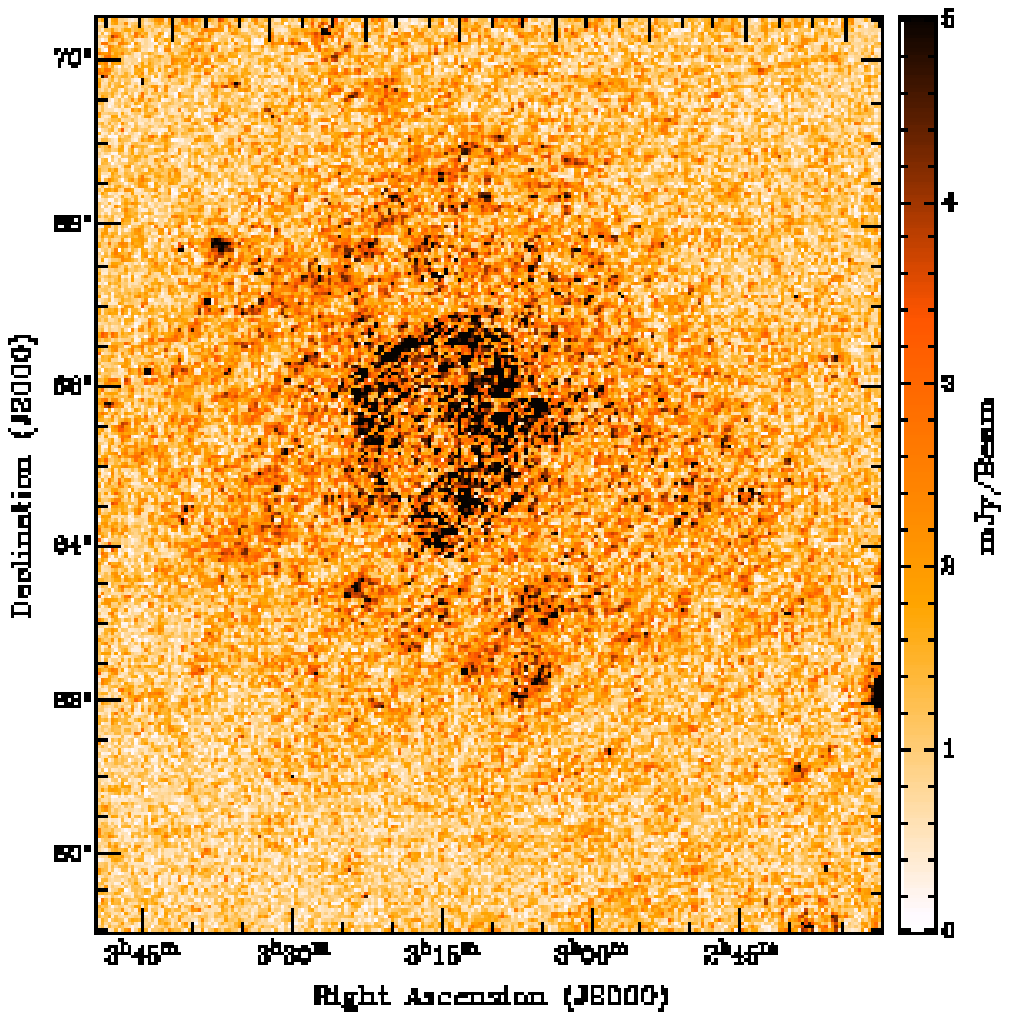}}
   \resizebox{0.9\columnwidth}{!}{\includegraphics{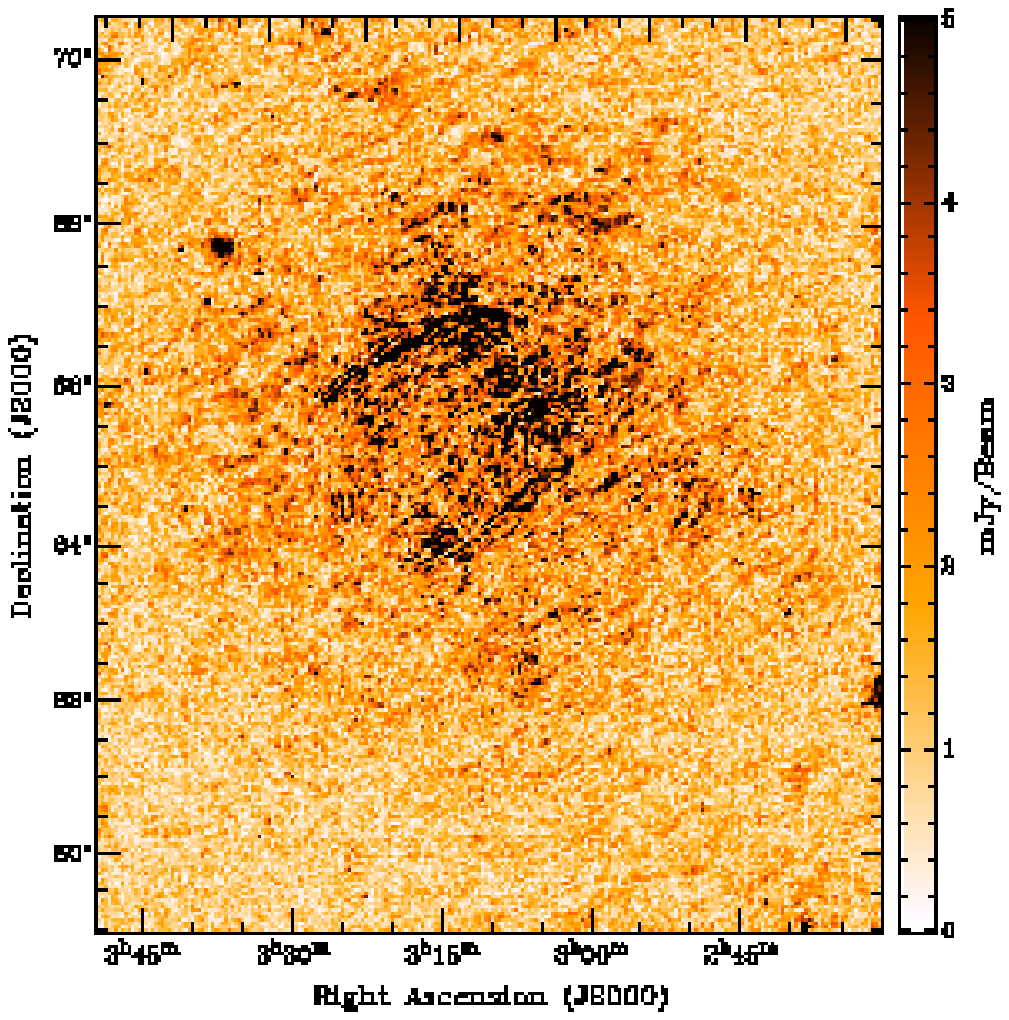}}
   
   \resizebox{0.9\columnwidth}{!}{\includegraphics{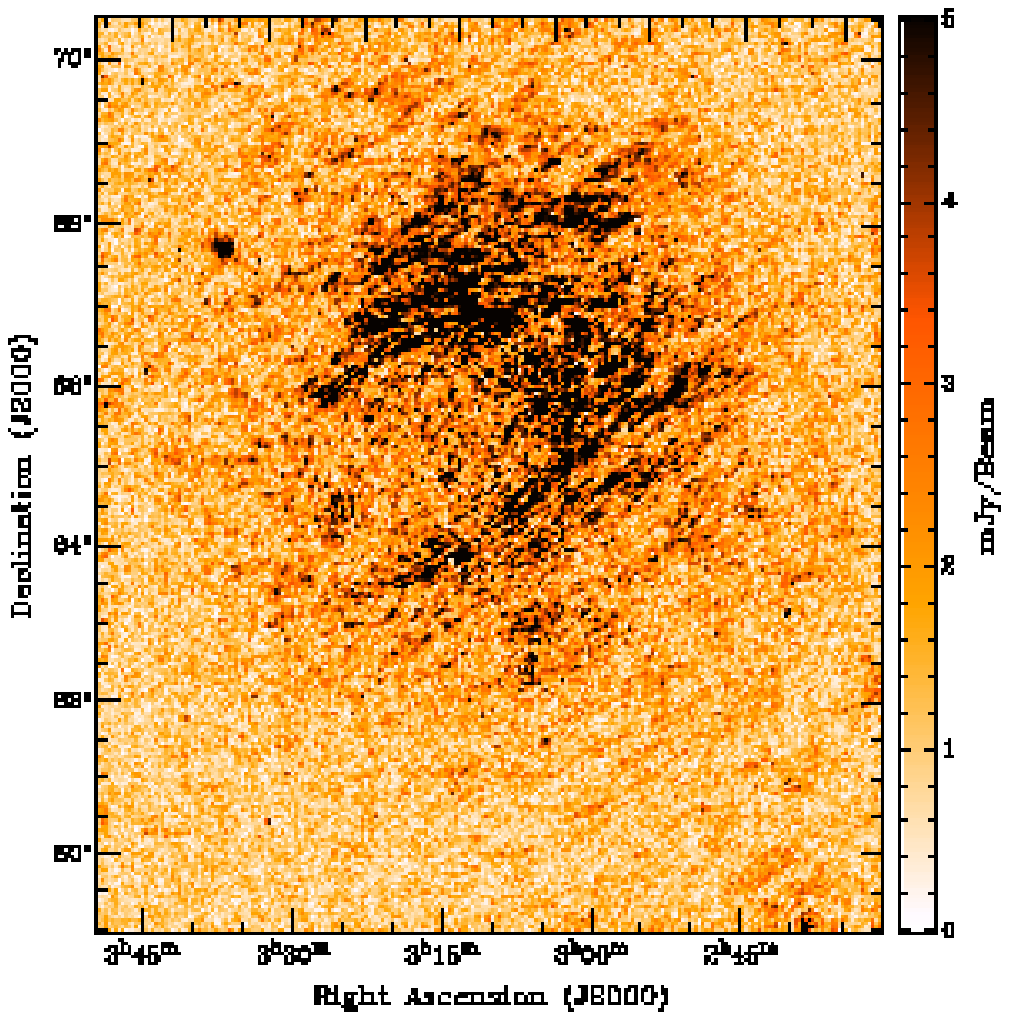}}
   \resizebox{0.9\columnwidth}{!}{\includegraphics{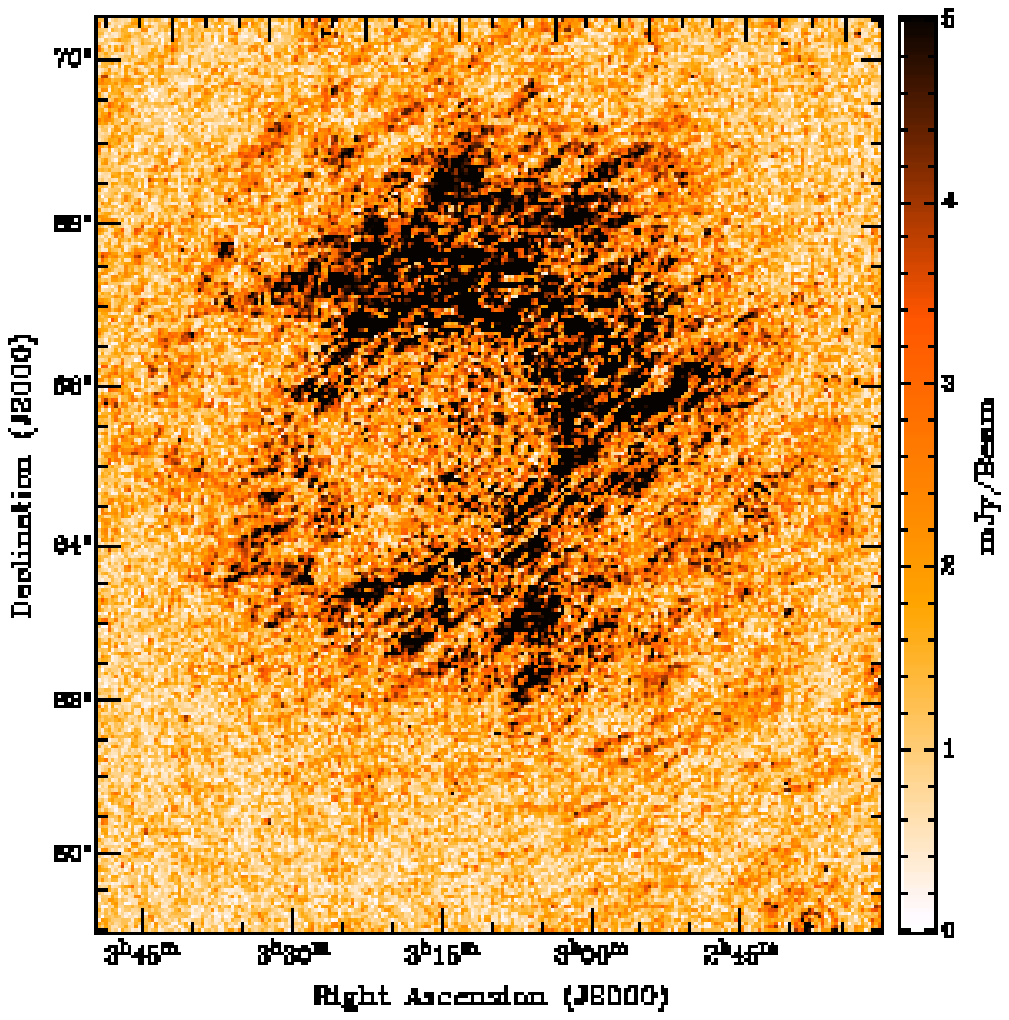}}
   
   \resizebox{0.9\columnwidth}{!}{\includegraphics{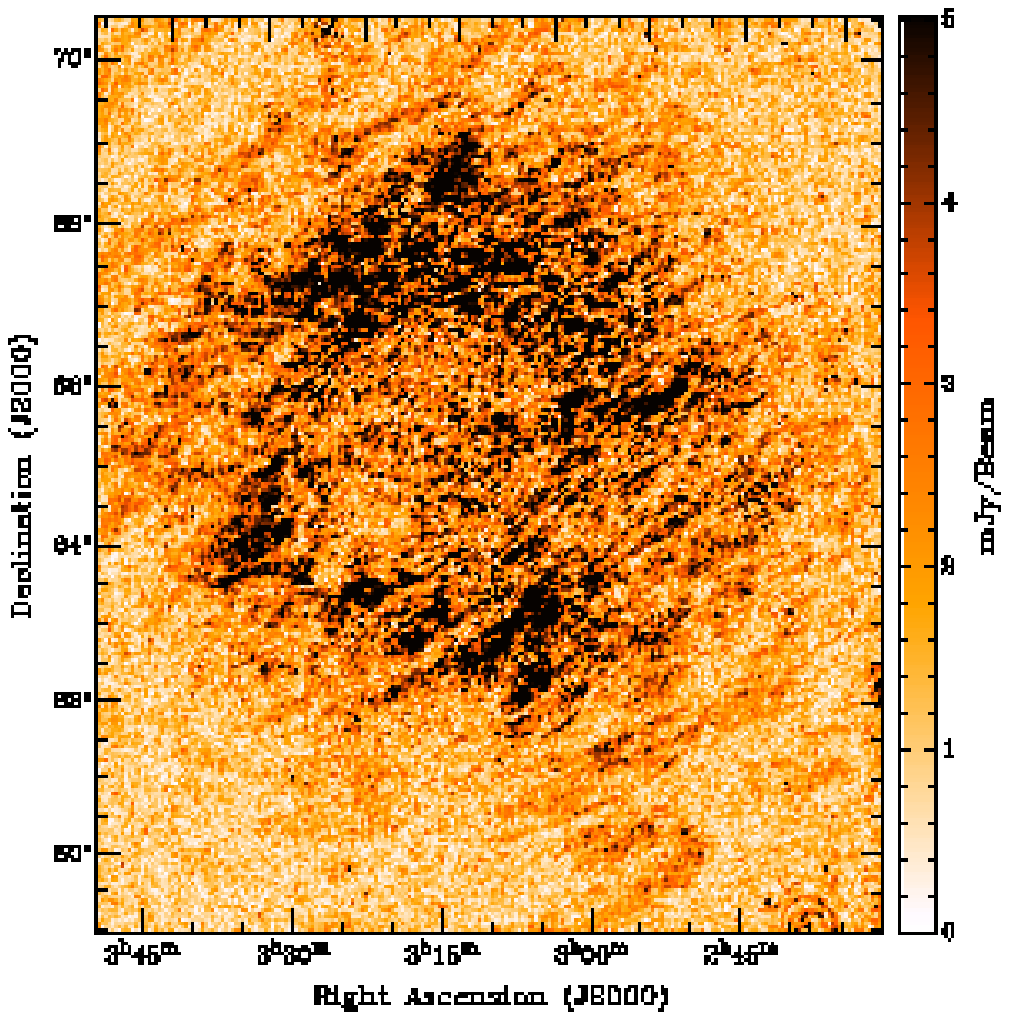}}
   \resizebox{0.9\columnwidth}{!}{\includegraphics{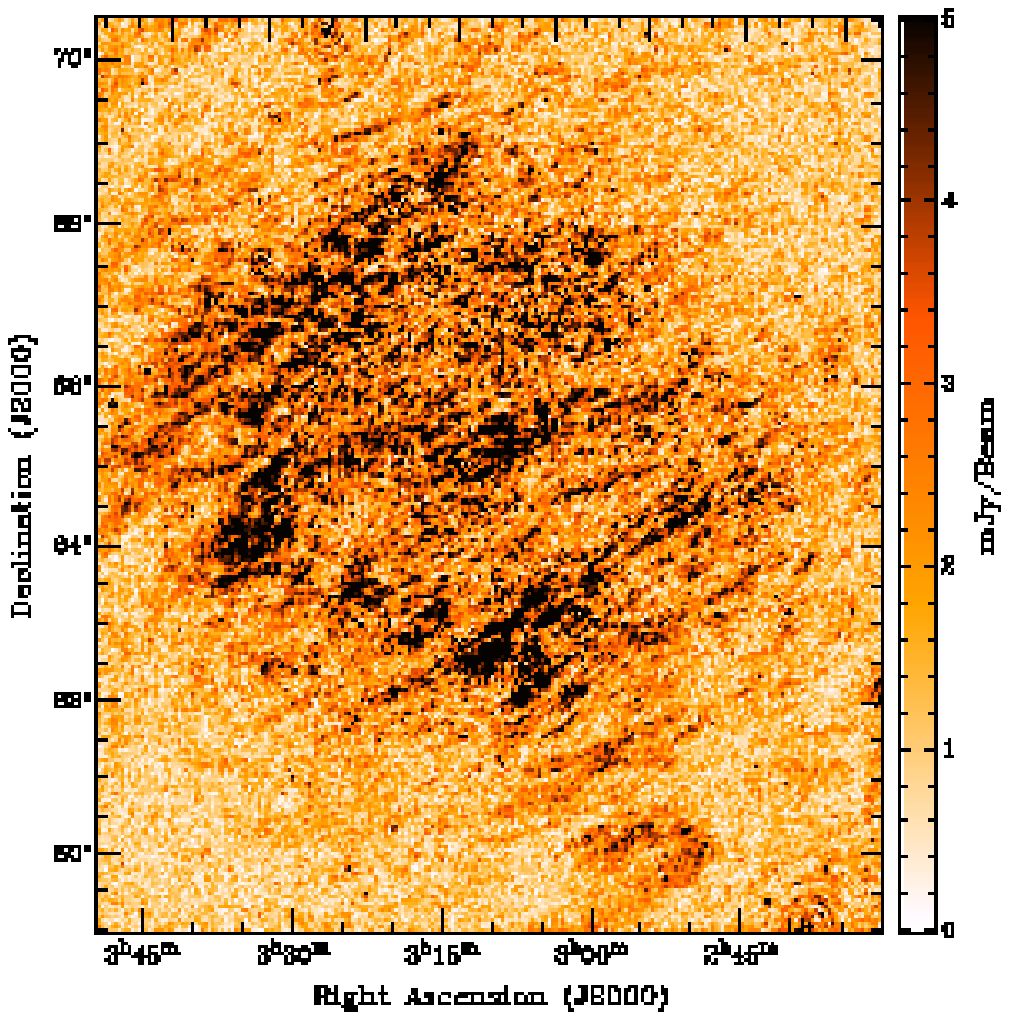}}
   \caption{Frames of the RM cube at different values: RM$ = -5$~rad~m$^{-2}$ (top left), RM$ = -4$~rad~m$^{-2}$ (top right),  RM$ = -3$~rad~m$^{-2}$ (middle left), RM$ = -2$~rad~m$^{-2}$ (middle right), RM$ = -1$~rad~m$^{-2}$ (bottom left), RM$ = 0$~rad~m$^{-2}$ (bottom right). The conversion factor is 1~mJy~beam$^{-1}$ = 1~K.}
              \label{RM_cube}
    \end{figure*}

For the output RM cube, we selected the RM interval $[-100,100]$~rad~m$^{-2}$ where the Galactic emission most likely appears. From the RM synthesis cube we found that all the polarized emission is concentrated in the interval [-7,1]~rad~m$^{-2}$, which means that the line--of--sight component of the Galactic magnetic field is mainly pointed away from the observer. Figure~\ref{RM_cube} shows the most significant frames of the RM cube. 
%
%
Most of the frames still show unsubtracted polarized point sources at the level of a few mJy. In particular they appear to be present at the edge of the field, where the instrumental polarization is stronger. 

The polarized emission shows a very peculiar trend as a function of Faraday depth. At $-5$~rad~m$^{-2}$ the emission is located in the center of the image, describing a bubble with a diameter of  $\sim 2^\circ$. The frame at $-3$~rad~m$^{-2}$ looks completely different: in the centre of the image, at the location of the peak of the intensity, there is now a hole in the emission, surrounded almost completely by diffuse polarization. This pattern becomes even more evident at $-2$~rad~m$^{-2}$ before disappearing again at RM = 0. This frame shows the peak of the polarized emission together with the greatest spread throughout the field. There is an evident orientation in the spatial pattern with `stripes' of emission running from South-West to North-East almost parallel to lines of constant Galactic latitude. This pattern was already apparent in the 350~MHz data in HKB.

\begin{figure}
   \centering
   \resizebox{0.9\columnwidth}{!}{\includegraphics{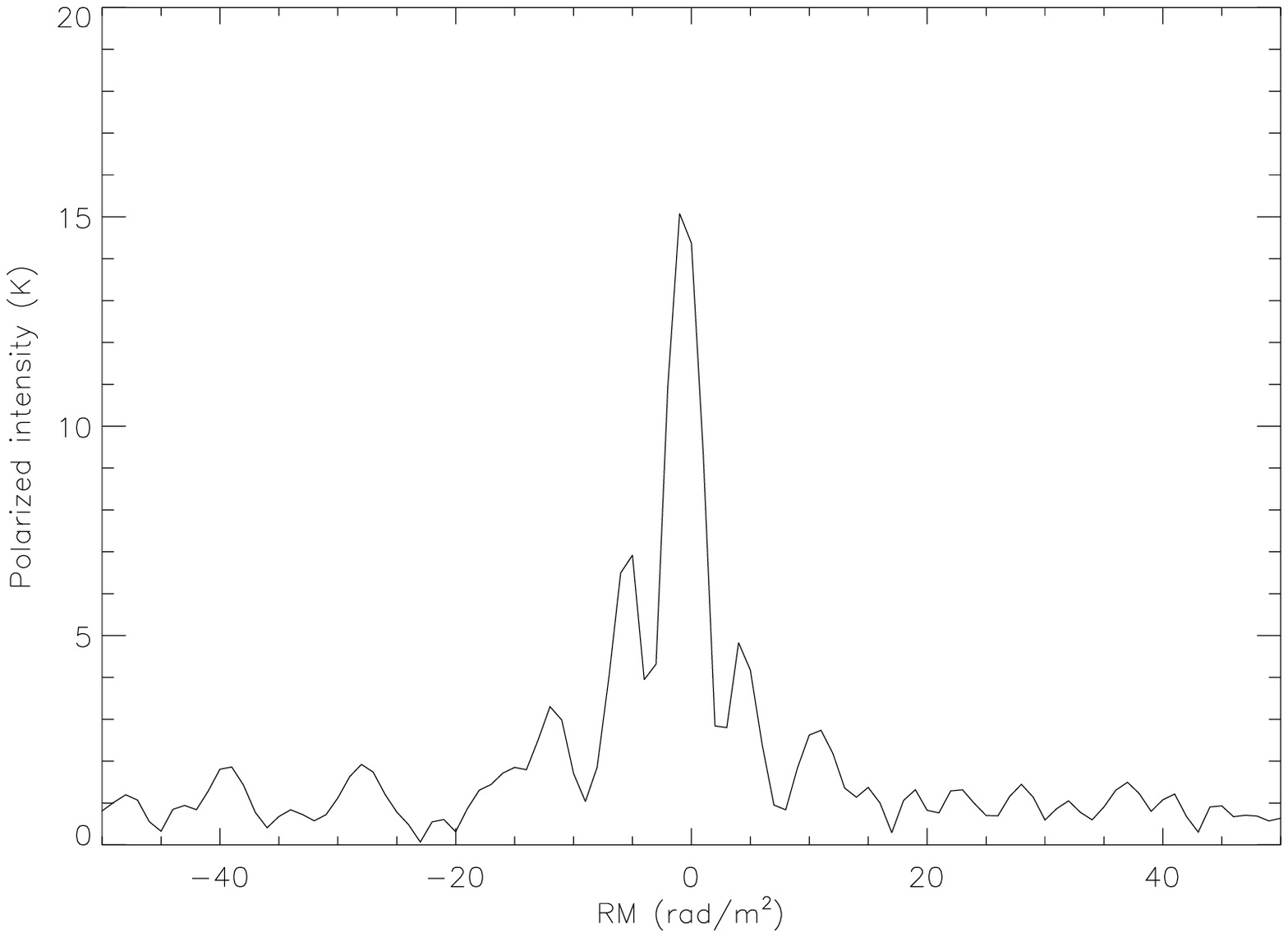}}
   \resizebox{0.9\columnwidth}{!}{\includegraphics{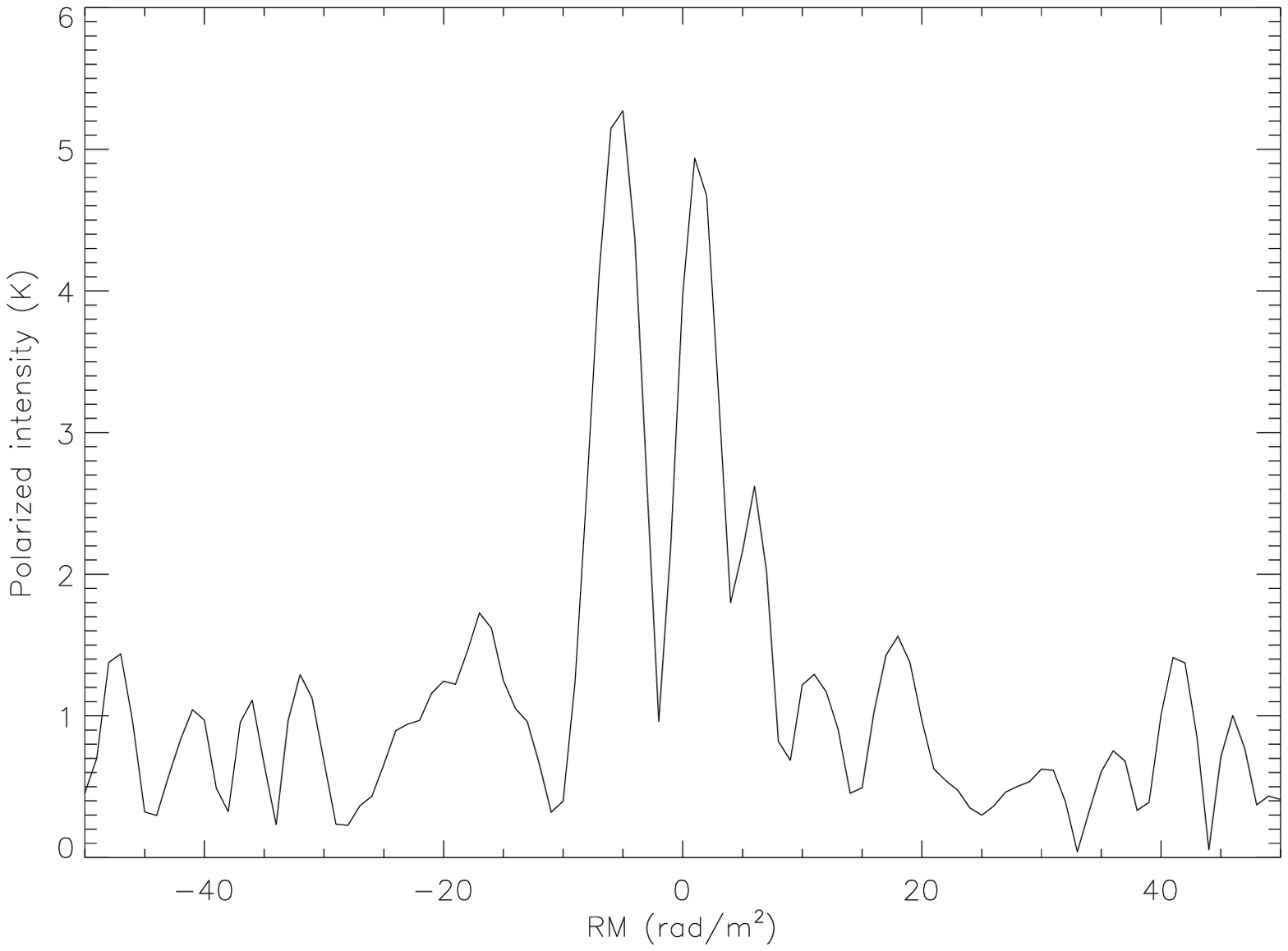}}
   \resizebox{0.9\columnwidth}{!}{\includegraphics{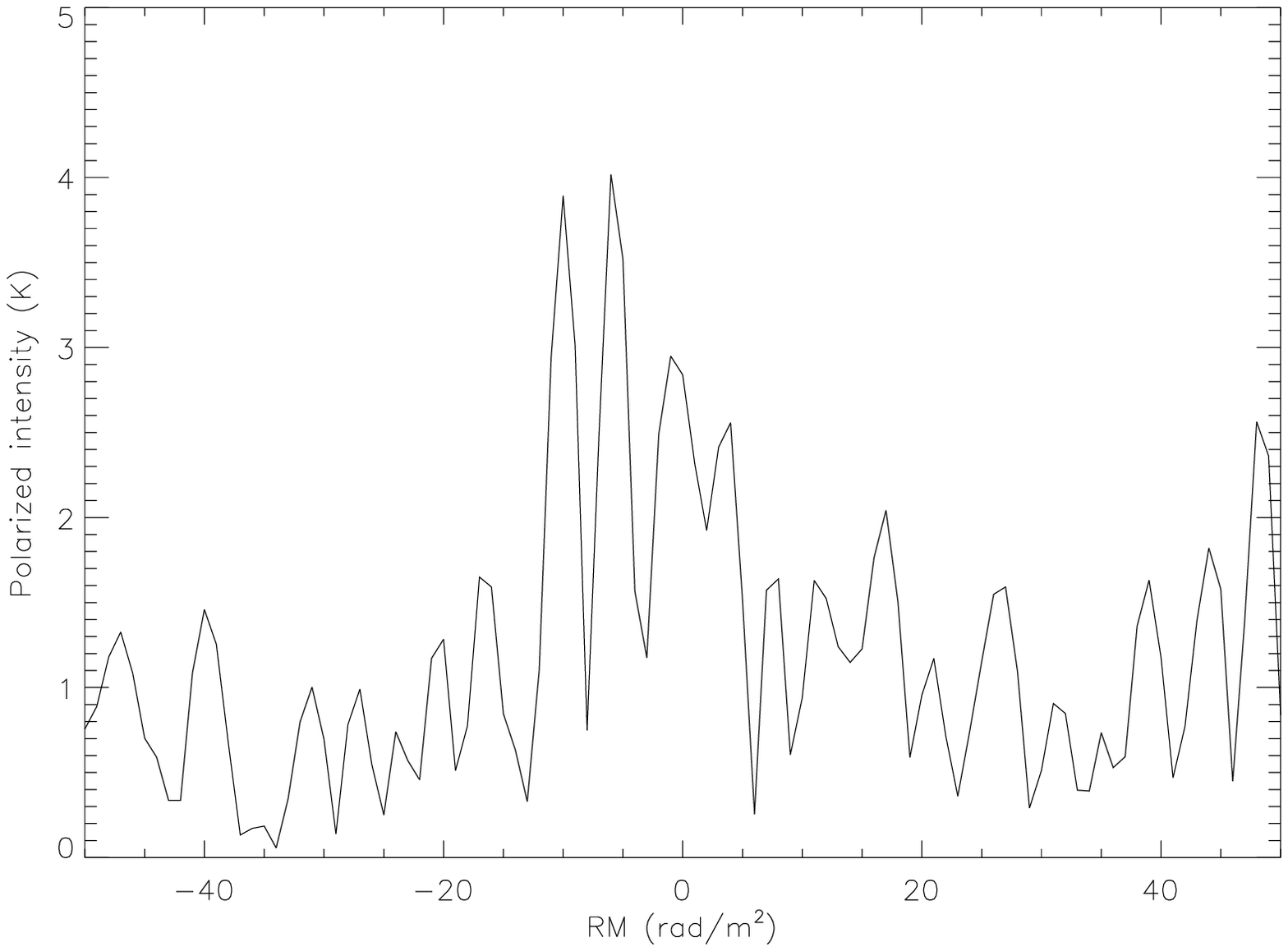}}
   \resizebox{0.9\columnwidth}{!}{\includegraphics{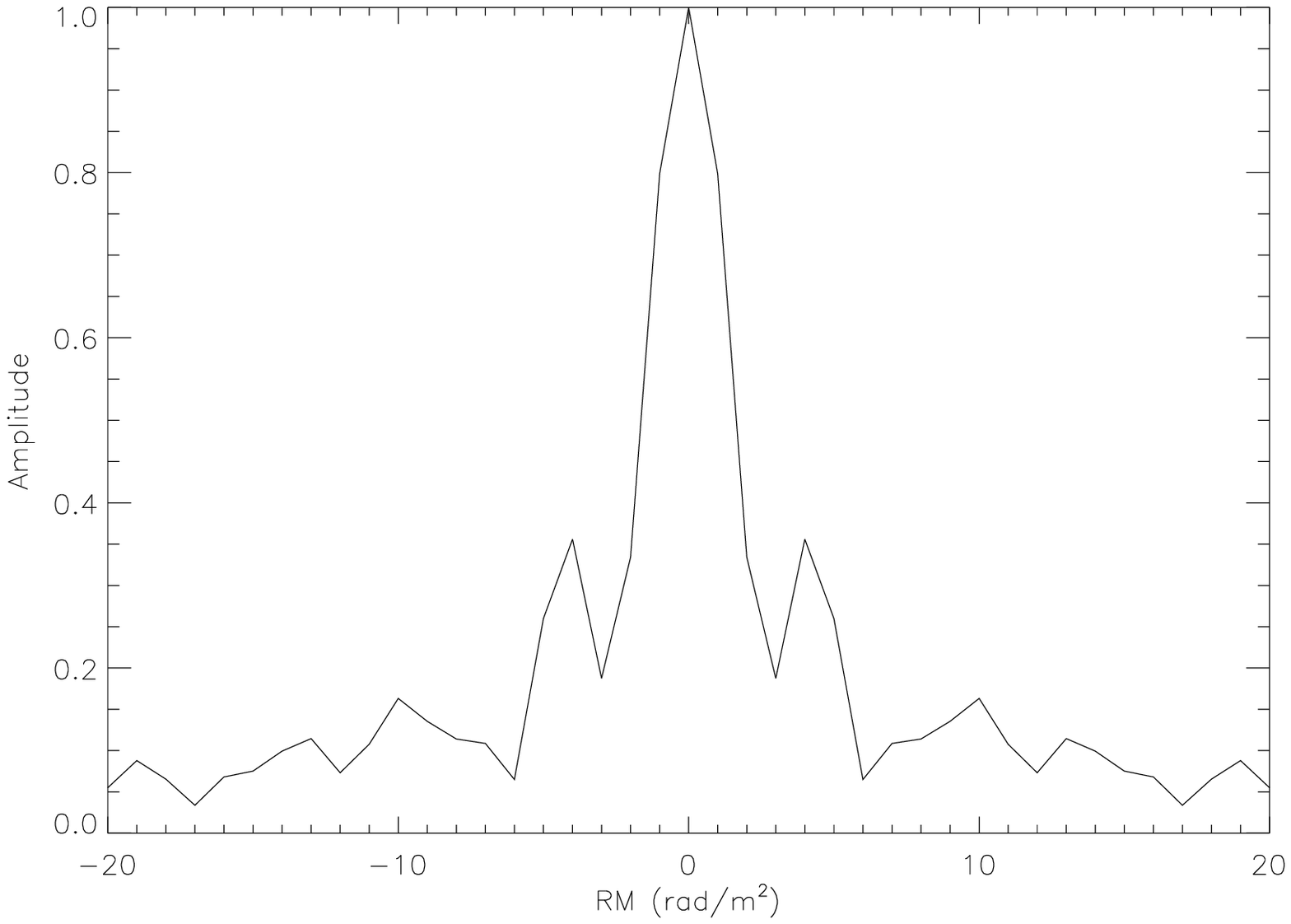}}
   \caption{Examples of Faraday spectra along different lines of sight together with the RMSF (bottom figure) of our observations.}
              \label{far_depth_lines}
    \end{figure}

In Figure~\ref{far_depth_lines} we show a few examples of polarized emission as a function of Faraday depth which further show the complexity of the interstellar medium through various lines of sight. In the first panel a bright point source lies at RM = 0. The slight asymmetry of the first side lobe might already indicate that the source is slightly more complicated than it looks. The second panel shows a double source: two peaks of the same intensity appear at -5~rad~m$^{-2}$ and 1~rad~m$^{-2}$. The third panel shows an even more complicated profile where multiple peaks are present at different RM values. 
%
%

\begin{figure*}
   \centering
   \resizebox{1.0\hsize}{!}{\includegraphics{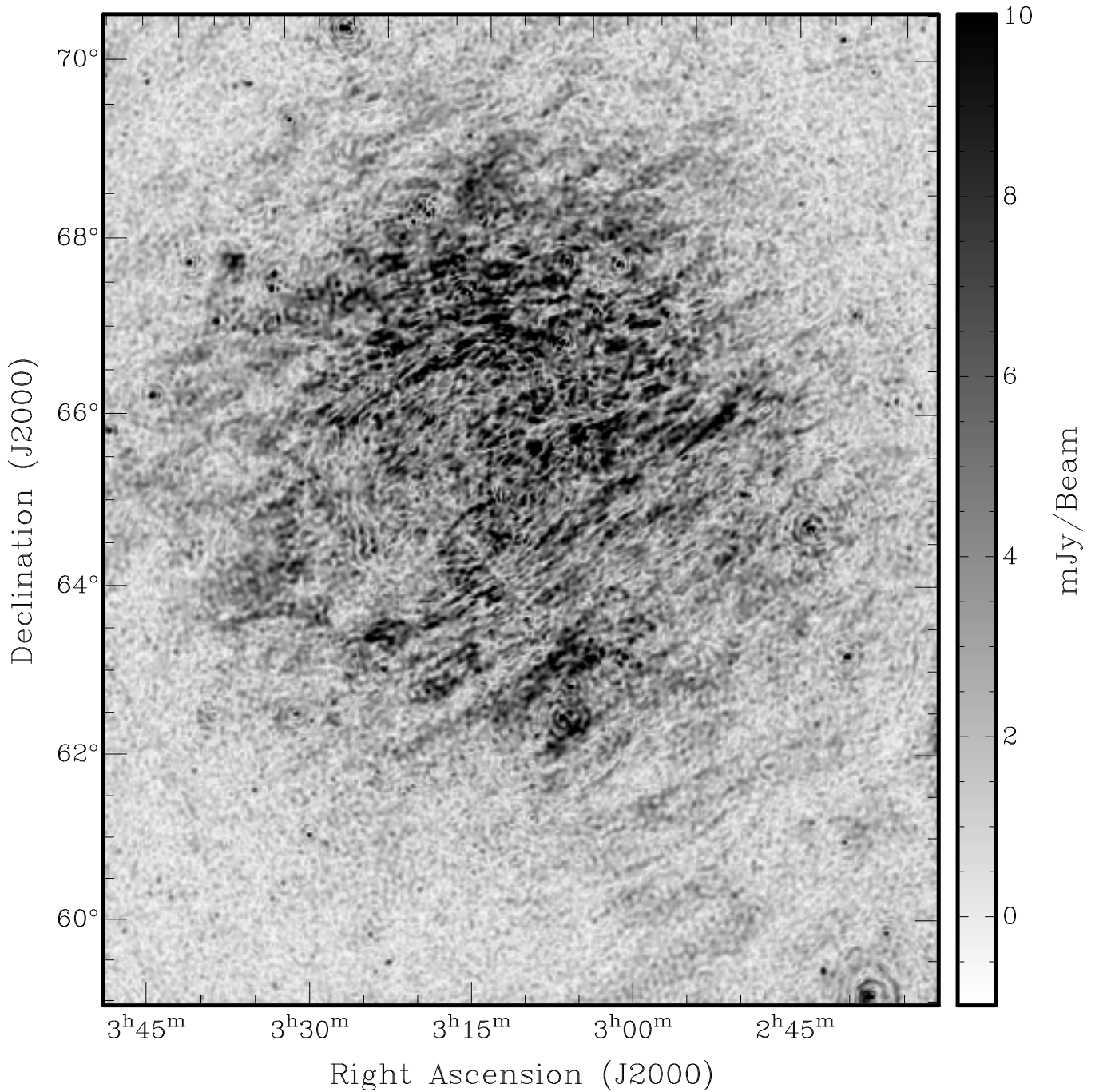}}
   \caption{Image of the total polarized intensity integrated along the Faraday depth. The conversion factor is 1~mJy~beam$^{-1}$ = 1~K.}
   \label{total_polarization}
   \end{figure*}

It is possible to compute the total polarized intensity $P$ by integrating along the Faraday depth. According to Brentjens (2007), it can be computed as follows:
\begin{eqnarray}
    P = B^{-1} \sum_{i=-7}^0 \left( P_{i} - \sigma_P \sqrt{\frac{\pi}{2}} \right)
\end{eqnarray}
where $\sigma_P \sim 1.4$~K is the noise in polarization, $P_i$ is the polarized map at the RM value $i$ and the sum is over the frames of the RM cube which contain relevant emission. The factor $B$ represents the area of the restoring beam divided by the interval between two frames of the RM cube. In our case we do not deconvolve the RM cube so we use the Gaussian which best fits the peak of the dirty beam ignoring the side lobe contribution. This approximation is justified by the fact that most of the polarized flux in the individual frames of the RM cube is about 7~K or lower, so its side lobe noise is lower than the thermal noise. Figure~\ref{total_polarization} shows the integrated polarized intensity.

Considering the inner $6^\circ \times 6^\circ$ square where we corrected for the primary beam shape, the $rms$ of polarization fluctuations is $7.2 \pm 0.8$~K.
%
%

\subsection{Comparison with the 350~MHz WSRT results and the nature of the `ring'}
\label{comp_350}

We compared our polarization results with the 350~MHz data which have similar angular resolution. The most striking feature present in the 350~MHz data is a ring-like structure which was first detected at 408~MHz in the late sixties (Bingham \& Shakeshaft 1967) and then observed again at 21~cm with the Green Bank telescope (Verschuur 1969). 

In the 350~MHz data, the ring is mostly visible in polarization angle. HKB did not perform RM synthesis but determined the rotation measures via the traditional linear fit to $\lambda^2$. They found that the RM values corresponding to the ring are mostly in the interval $-8 <$ RM $< -4$~rad~m$^{-2}$ and find positive RM values -- up to 12~rad~m$^{-2}$ -- in the North-East and South-West corners of the field.

In their work they concluded that the ring is probably a magnetic feature in the field parallel to the line of sight. They found no evidence of total intensity emission associated with the detected polarized structure.

This ring-like structure is still present at 2~m wavelength, centred at $\alpha \sim 3^{\rm h} 15^{\rm m}$, $\delta \sim 65^\circ$ and seen in emission for RM~$ < -4$~rad~m$^{-2}$ and as a lack of emission for $-3 <$ RM $< -1$~rad~m$^{-2}$. 

\begin{figure*}
   \centering
   \resizebox{\columnwidth}{!}{\includegraphics{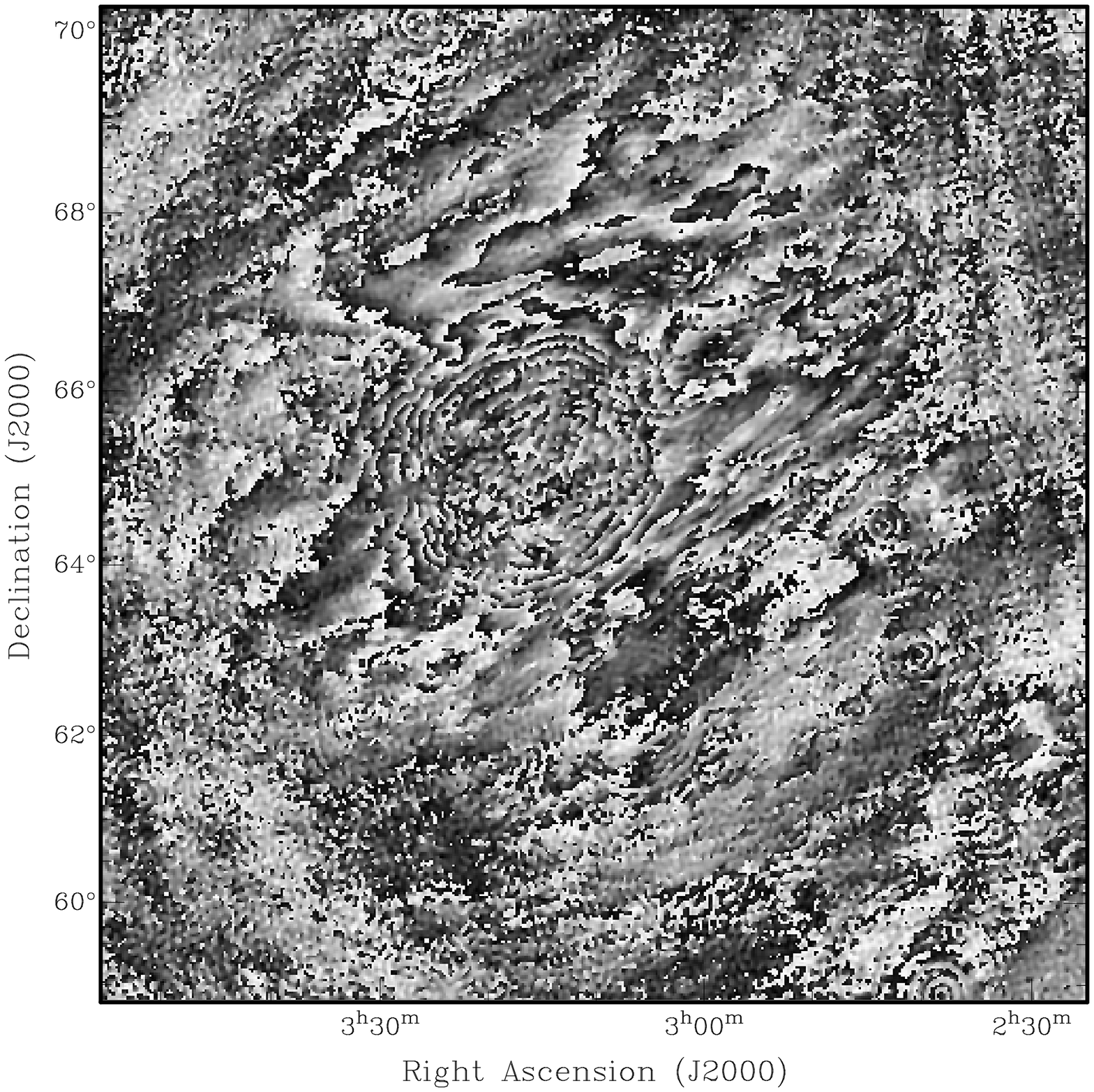}}
   \resizebox{\columnwidth}{!}{\includegraphics{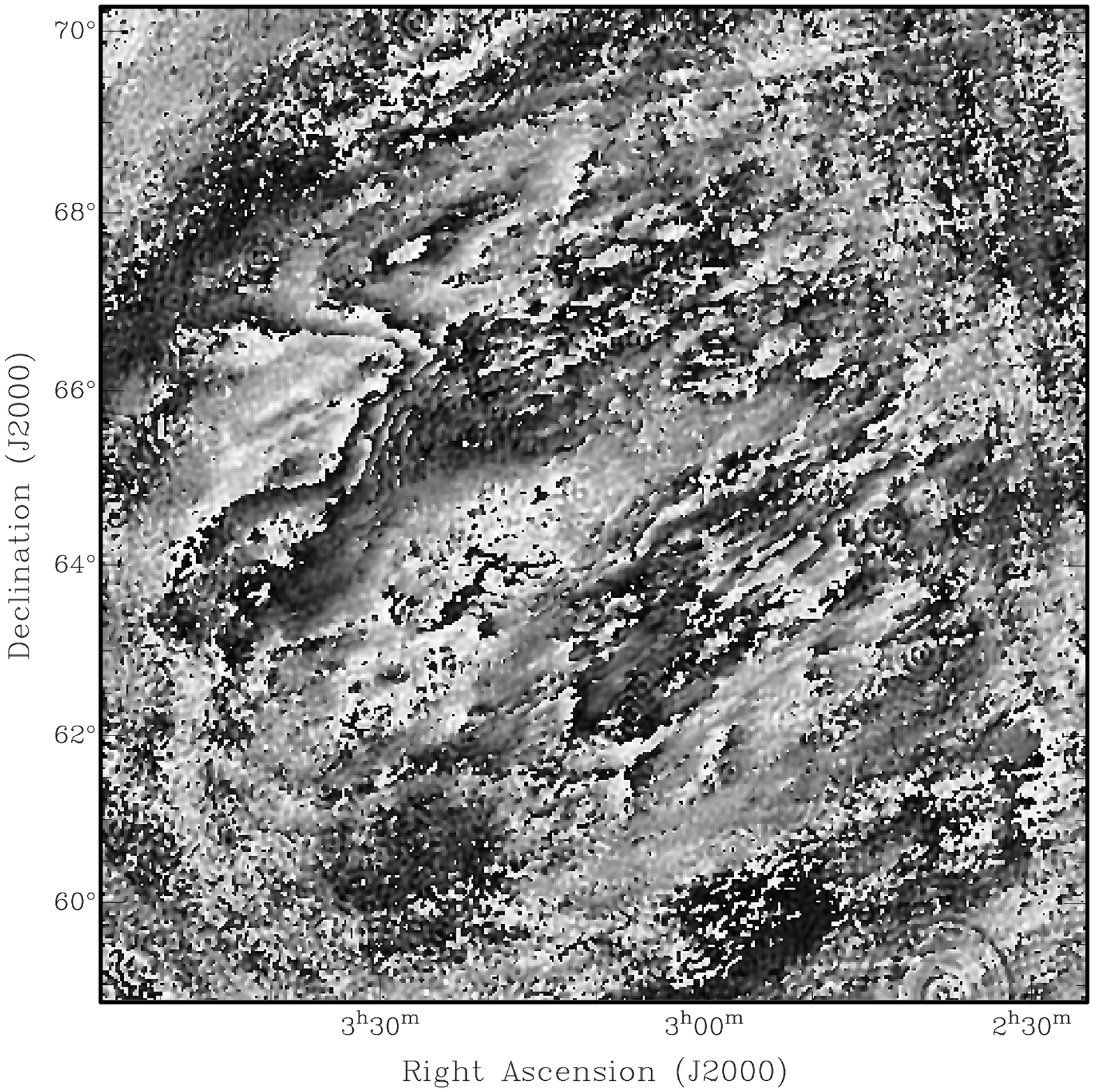}}
   \caption{Images of polarization angles for the frames at RM $ = -2$~rad~m$^{-2}$ (left) and RM $ = 0$~rad~m$^{-2}$ (right). In the color scale, black corresponds to $-90^\circ$ and white to $+90^\circ$.}
              \label{pol_ang_maps}
    \end{figure*}

The map of polarization angles shown in Figure~\ref{pol_ang_maps} makes the ring more evident.
We can observe two regimes in the spatial distribution of polarization angles. Outside the ring, polarization angles are uniform on scales of several tens of arcmin, whereas there are variations in the polarization angle on much smaller scales inside it. At 350~MHz instead, polarization angles show structures on bigger scales inside the ring.

The structure that we observe as a ring could therefore be due to a change in the spatial gradient in the polarization angle and this gradient could also be responsible for depolarizing the signal inside the ring at $-3 <$~RM~$ < -1$~rad~m$^{-2}$ where the polarized emission disappears. For the rest, the large scale distribution in polarization angles displays the same orientation as the polarized intensity, from South-West to North-East. 

There are two differences between the 150~MHz and the 350~MHz data. The 2~m data show almost no positive RM values. This discrepancy might be due to the fact that the 350~MHz data are mosaic observations and, therefore, could lack the large scale emission, although HKB reject this possibility. A missing large scale structure can generate changes in the gradient of the spatial distribution of RMs (Schnitzeler et al. 2007).

The second important difference is the total intensity counterpart. Although there is no direct and strong correlation between total intensity and polarization, in Section~\ref{diff_gal} we showed that there are fluctuations in the Galactic radio background on scales between 13~arcmin and one degree. Assuming a spectral index $\beta=2.5$ for the synchrotron emission, these fluctuations scale down to $\sim 1.7$~K at 350~MHz. This value is still a factor $\sim$2 above their noise and should allow a partial detection. The cause of this discrepancy between the 150~MHz and 350~MHz data remains unclear.

In light of the picture we have outlined so far, we suggest an origin for the nature of the ring different from the explanation proposed by HKB. They interpreted the ring as a magnetic structure, like a tube, which generates the variations of the magnetic field along the line of sight responsible for the variations in the observed RM. 

At 150~MHz we observe diffuse total intensity emission and, therefore, we detect fluctuations in the perpendicular component of the magnetic field, responsible for the synchrotron emissivity. We do not, however, observe a tight correlation between the polarized and total intensity components. It is hard to imagine a mechanism which generates fluctuations only in the parallel component of the magnetic field, leaving the perpendicular one unperturbed.

\begin{figure}
   \centering
   \resizebox{\columnwidth}{!}{\includegraphics{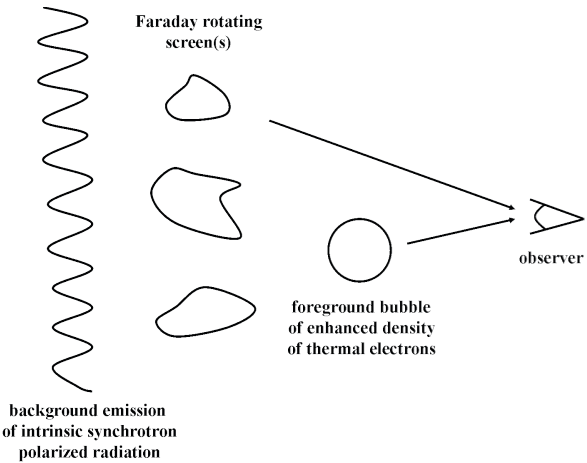}}
   \caption{A possible picture of the line-of-sight magnetized interstellar medium in the Fan area. The intrinsically polarized background emission passes through multiple Faraday screens which make the spatial distribution of the polarized emission more clumpy. These screens generate the bulk of the polarized emission at RM~$= [-2,1]$~rad~m$^{-2}$. A bubble of ionized thermal plasma sits between the Faraday screens and the observer. This bubble further rotates the polarized emission pushing it to RM~$\sim$-5~rad~m$^{-2}$.}
   \label{cartoon_pol}
   \end{figure}

What is easier to imagine is a picture where at least two screens are acting in front of background synchrotron emission. The background emission is Faraday rotated by a large scale screen which originates the bulk of the emission at RM~$= [-2,1]$~rad~m$^{-2}$. In front of this, another perturbation in the distribution of the thermal electrons along the line of sight, like a `bubble', would further rotate the polarized intensity. Figure~\ref{cartoon_pol} sketches this scenario.
 
This possible explanation would be supported by the morphology observed in the frames of the RM cube. As we mentioned above, the frames at -3~rad~m$^{-2}$ and -2~rad~m$^{-2}$ show a cavity where the frame at -5~rad~m$^{-2}$ shows emission. If a foreground bubble with higher electron density was present in front of polarized background emission, it could just shift the polarized emission to higher RM values, as we observe.

This only represents a qualitative interpretation of the data, and quantitative simulations that can predict the observed distribution of polarized emission as a function of Faraday depth in the interstellar medium are needed in order to obtain a more constraining picture.
%
%
%
%

\section{Power spectrum analysis}
\label{power_spec}

In Section~\ref{diff_gal} and Section~\ref{pol_em} we have described fluctuations in the diffuse emission through the $rms$ at a certain angular scale. We also argued that the total intensity fluctuations at arcmin scales are affected by source confusion and, therefore, not fully representative of the diffuse emission alone. In this section we present a power spectrum analysis in order to characterize completely the emission as a function of the angular scale.

The angular power spectrum is commonly used in cosmology to define the spatial properties of diffuse radiation. In interferometric observations, the power spectrum is usually computed in the $uv$ plane, because the interferometer directly measures the Fourier components of the brightness temperature of the sky (White et al. 1999; Myers et al. 2003; Morales \& Hewitt 2004). 

However, in this paper we compute the power spectra from the image as this approach allows for two simplifications. First, the correction for the primary beam of the interferometer can be applied much more easily in the image plane than in the $uv$ plane. In the image plane the primary beam represents a simple multiplicative effect whereas it is a convolution in the $uv$ plane (Thompson et al. 1986). Second, the total polarization integrated along the whole Faraday depth is more easily computed in the image plane.

Given this, we estimated the angular power spectrum $C^X_\ell$ as (Seljak 1997):
\begin{eqnarray}
   C^X_\ell = \left\{ \frac{\Omega}{N_\ell} \sum_{\bf l} X({\bf l}) X^*({\bf l}) - \frac{\Omega \sigma^2_{\rm{noise}}}{N_b} \right\} b^{-2}(\ell)
	\label{pow_spec_def}
\end{eqnarray}

where $X$ indicates either the total intensity $I$ or the polarized intensity 
$P$, $\ell=\frac{180}{\Theta}$ where $\Theta$ is the angular scale in degrees, $\Omega$ is the solid angle in radians, ${N_\ell}$ is the number of Fourier modes around a certain $\ell$ value, $\bf{X}$ and $\bf{X}^*$ are the Fourier transform of the image and its complex conjugate respectively, $\bf{l}$ is the two dimensional coordinate in Fourier space, $\sigma_{\rm{noise}}$ is the $rms$ noise and $N_b$ is the number of independent synthesized beams in the map. The factor $b^{2}(\ell)$ is the power spectrum of the window function (Tegmark 1997). Since interferometric images represent the true sky brightness convolved with the dirty beam (the Fourier transform of the weighted $uv$ coverage), in our case the factor $b^{2}(\ell)$ is the power spectrum of the dirty beam.

The number of modes around a certain $\ell$ value depends on the bin width in Fourier space and has a minimum dictated by the width of the field of view of the instrument. The 25~m dish of the WSRT telescope gives $\ell_{min} \sim 40$, therefore we choose $\Delta \ell = 50$ as the bin width for computing the power spectrum.

\begin{figure}
   \centering
   \resizebox{\columnwidth}{!}{\includegraphics{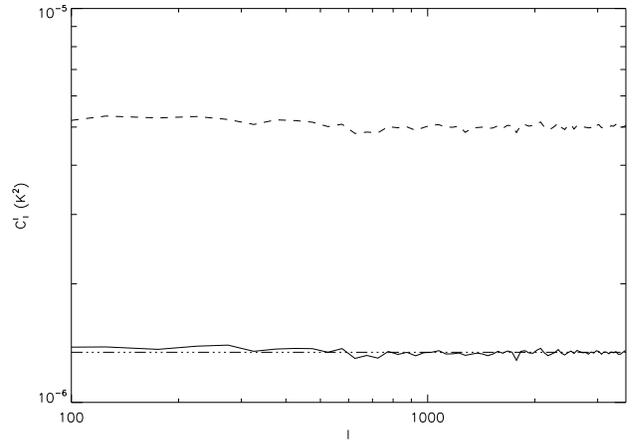}}
   \caption{Solid line: average power spectrum computed on a $6^\circ \times 6^\circ$ noise map (see text for details). Triple--dot--dashed line: expected power spectrum of the thermal noise. Dashed line: average power spectrum of the noise after the primary beam correction is applied.}
   \label{noise_pow_spec}
   \end{figure}

The $rms$ noise can be measured directly from the map, as we did in Section~\ref{diff_gal}. However, since we have corrected the map for the primary beam, the noise was affected by the correction as well. In order to understand how the expected noise power spectrum changed under the primary beam correction, we generated 100 realizations of a $6^\circ \times 6^\circ$ noise map where the pixel values were drawn from a Gaussian distribution with $\sigma_{\rm{noise}} = 0.75$~mJy~beam$^{-1}$ on 2~arcmin scales. Afterwards we multiplied the map by the primary beam shape. We computed a power spectrum for each noise map, and for each noise map after correcting for the primary beam shape. The resulting average power spectra are shown in Figure~\ref{noise_pow_spec}.
%
%

We can see that the power spectrum remains flat at all the angular scales even after the map is multiplied by the primary beam function, but the amplitude increases. We find that the ratio between the two spectra is $\sim$3.7, therefore we account for this factor in subtracting the $rms$ noise in Equation~\ref{pow_spec_def}.

Finally, it is relevant to note that Equation~\ref{pow_spec_def} relates the power spectrum and the $rms$ fluctuations $T^X_{rms}$ measured from a map as:
\begin{eqnarray}
   T^X_{rms}= \sqrt{\sum_{N_{\rm bin}} \Delta \ell \, \frac{\ell \, C^X_\ell \, b^2(\ell)}{2 \pi}}
\end{eqnarray}
where $N_{\rm bin}$ is the number of bins used to compute the power spectrum.

\begin{figure}
   \centering
   \resizebox{\columnwidth}{!}{\includegraphics{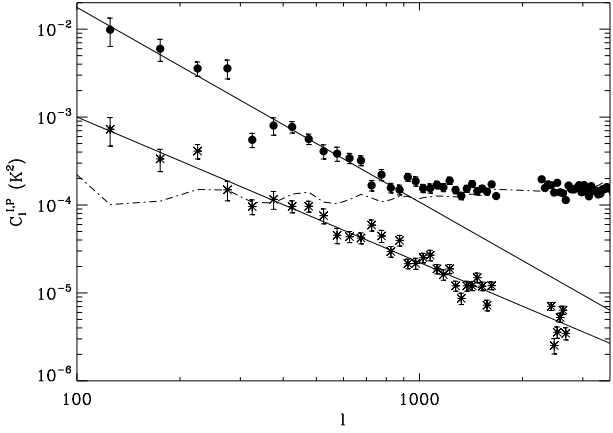}}
   \caption{Filled circles: total intensity power spectrum from the residual image of Figure~\ref{full_stokes} with the best power--law fit superimposed (solid line). Asterisks: polarization power spectrum from the total polarized intensity map of Figure~\ref{total_polarization} with the best power--law fit superimposed (solid line). Dot--dashed line: power spectrum due to residual point sources (see text for details). The plotted 1~$\sigma$ error bars only account for the statistical errors. Power spectra are computed in the inner $6^\circ \times 6^\circ$ square of the map.}
\label{fan_pow_spec}
    \end{figure}

Figure~\ref{fan_pow_spec} presents the power spectrum down to 3~arcmin, where we avoided the smallest scales which are the most affected by residual calibration and ionospheric errors as we noted in Section~\ref{diff_gal}.
%

The total intensity power spectrum shows two clearly different behaviours as a function of the angular scale. At large angular scales there is a power--law behaviour, with power decreasing at increasing angular scale. This behaviour is typical of the Galactic diffuse emission observed at higher frequencies and higher angular resolutions (Bennett et al. 2003, La Porta et al. 2008). 

Around $\ell \sim 900$ the power spectrum flattens and remains flat down to 3~arcmin scales. This behaviour is expected if point sources dominate the emission. 

We have found in Section~\ref{diff_gal} that the image made by discarding the short baselines contains no diffuse emission, therefore we expect the power spectrum to be dominated by point sources at high $\ell$ values.

The angular power spectrum due to unsubtracted point sources is expected to be flat if the point sources have a random spatial distribution (Tegmark \& Efstathiou 1996), therefore it can be modelled as an additional noise term once its $rms$ is known. We used the $\sigma_{\rm{ps}}$ value measured from the map of Figure~\ref{fan_point_sources} as representative of the $rms$ fluctuations due to unsubtracted point sources and estimated the corresponding angular power spectrum as:
\begin{eqnarray}
       C_{\ell}^{\rm{ps}} = \frac{\Omega \, \sigma^2_{\rm{ps}}}{N_b \, b^{2}(\ell)}.
\label{pow_spec_ps}
\end{eqnarray}

In Figure~\ref{fan_pow_spec} the power spectrum expected from residual point sources agrees very well with the power spectrum computed from the map for $\ell > 900$, corresponding to $\sim$12~arcmin. We can therefore conclude that the power spectrum down to $\ell \sim 900$ is representative of the Galactic diffuse foreground emission whereas it is dominated by residual point sources at higher $\ell$ values. The sensitivity to diffuse structure on small angular scales is therefore not limited by instrumental noise, but by source confusion. The analysis presented in Section~\ref{diff_gal} fully agrees with the power spectrum analysis.

The power spectrum of the total intensity diffuse emission was fitted by a power--law down to $\ell = 900$: 
\begin{eqnarray}
    C_{\ell}^I = C_{400}^{I} \left({\ell \over 400}\right)^{\beta^I_\ell}. 
\end{eqnarray}
The best fit values are $C_{400}^{I} = 0.0019 \pm 0.0003$~K$^2$ and $\beta^I_\ell = -2.2 \pm 0.3$. The slope is within the range of slopes [-2,-3] found by all the previous measurements at higher frequencies (Tegmark \& Efstathiou 1996, Giardino et al. 2001, Bennett et al. 2003, La Porta et al. 2008).

The polarization power spectrum is fainter than the total intensity one and shows a regular decrease from the largest scales down to 4~arcmin scales. 
The power spectrum analysis confirms that polarization data are not limited by confusion noise because of the very few polarized point sources, as the RM synthesis analysis already showed. The $rms$ value down to 4~arcmin therefore represents fluctuations in the diffuse Galactic polarized foreground.

Also the polarization power spectrum was fitted by a power--law down to $\ell = 2700$:
\begin{eqnarray}
    C_{\ell}^P = C_{700}^{P} \left({\ell \over 700}\right)^{\beta^P_\ell}. 
\end{eqnarray}
The best fit values are $C_{700}^{P} = 90 \pm 7$~(mK)$^2$ and $\beta^P_\ell = -1.65 \pm 0.15$. We note that the slope of the polarization power spectrum is flatter than the slope of the total intensity as was found by most of the studies of the synchrotron emission as a foreground of the cosmic microwave background (Tucci et al. 2000; Baccigalupi et al 2001; Bruscoli et al. 2002; Tucci et al. 2002).

\section{Discussion and conclusions}
\label{concl}

We have presented results from Westerbork observations of the Fan region at 150~MHz both in total intensity and in polarization, mainly focused on studying the foregrounds for the cosmological 21~cm line. Our observations are the deepest available so far, reaching a thermal noise of 0.75~mJy~beam$^{-1}$. However, on arcmin scales the main limitation comes from the classical confusion noise, which is $\sim$3~mJy~beam$^{-1}$ in the map.

For the first time total intensity and polarization fluctuations in the Galactic diffuse foreground emission have been measured at the frequencies and angular scales relevant for EoR experiments.

We have detected structure in the diffuse total intensity Galactic foreground at 150~MHz with an $rms$ fluctuation of 14~K at 13~arcmin resolution. The power spectrum analysis showed that the signal follows a power--law behaviour on the largest scales down to $\ell \sim 900$ where it becomes compatible with the expected contribution from residual point sources. 

The best power--law fit to the power spectrum of Galactic emission gave an amplitude of $C_{400}^{I} = 0.0019 \pm 0.0003$~K$^2$ at $\ell = 400$ and a slope $\beta^I_\ell = -2.2 \pm 0.3$.

We have used the RM synthesis technique to measure the polarized emission and we found that polarization is quite complex and structured both spatially and in Faraday depth. The RM synthesis technique has confirmed its status as a superb tool to detect multiple layers of emission at different Faraday depths.

Although the large scale distribution of polarized emission somewhat resembles the total intensity one, the lack of correlation on intermediate and small angular scales suggests that the polarized emission is generated through the Faraday screen mechanism. 

The `ring' already observed at 350 and 408~MHz has been detected again in our data. From the information available at different Faraday depths, we suggest that the ring is more likely to be a feature in the density of thermal electrons, like a foreground bubble, rather than a magnetic structure. 

The detection of polarization at 150~MHz indicates that the local medium must be Faraday thin, otherwise most of the polarization structures along the line of sight would vanish at 2~m due to Faraday depolarization. Some Faraday depolarization is also present in the Fan area, however, because we would expect $\sim$40~K of polarized signal on the basis of a direct extrapolation of the 350~MHz data. Instead we found 7.2~K of $rms$ fluctuations at 4~arcmin after integrating along the Faraday depth. 

The polarization angular power spectrum shows a monotonic decrease from the largest angular scales down to 4~arcmin. All the polarized signal comes from diffuse structure in the Galaxy. The best power--law fit to the power spectrum of polarized emission gave an amplitude of $C^P_{700} = 90 \pm 7$~(mK)$^2$ at $\ell = 700$ and a slope $\beta^P_\ell = -1.65 \pm 0.15$.

The fact that the Fan region shows a strongly polarized signal over scales ranging from several degrees down to 4~arcmin scales makes it a promising target to study the behaviour of the ionospheric Faraday rotation and to calibrate the instrumental polarization of the forthcoming low frequency arrays.

The results presented so far can be used to improve the predictions of the level of foreground contamination of the EoR signal.

The Fan region is obviously not a potential target of EoR observations, because of its proximity to the Galactic plane.
However, the power spectrum of total intensity fluctuations measured there can be extrapolated down to smaller angular scales and used to obtain an indication of the level of the fluctuations at higher Galactic latitude, where EoR observations will certainly be carried out. We will limit ourselves to presenting results in terms of two--dimensional power spectra which only describe the spatial properties of the sky brightness. The EoR signal has an intrinsic three dimensional nature which will ultimately be taken into account in order to remove the foregrounds and to characterize the statistics of the cosmological signal (Zaldarriaga et al. 2004, Datta et al. 2007, Iliev et al. 2008), but we postpone this analysis to future work.

In all the radio surveys of the Galaxy, the diffuse emission decreases away from the Galactic plane (see de Oliveira-Costa et al. 2008 for a recent analysis of multi-frequency all-sky data). In Figure~\ref{gal_plot} we plotted the behaviour of the $rms$ fluctuations as a function of Galactic latitude in the $2^\circ$ resolution all--sky map at 150~MHz and showed that the signal drops by a factor of ten already at moderate Galactic latitude, staying roughly constant up to the Galactic pole. Also in surveys at higher frequencies and with slightly better angular resolution (Jonas, Baart \& Nicolson 1998; Gold et al. 2008), it is possible to find ``cold" spots where the signal is ten times fainter than in the proximity of the plane or even lower. 

\begin{figure}
   \centering
   \resizebox{\columnwidth}{!}{\includegraphics{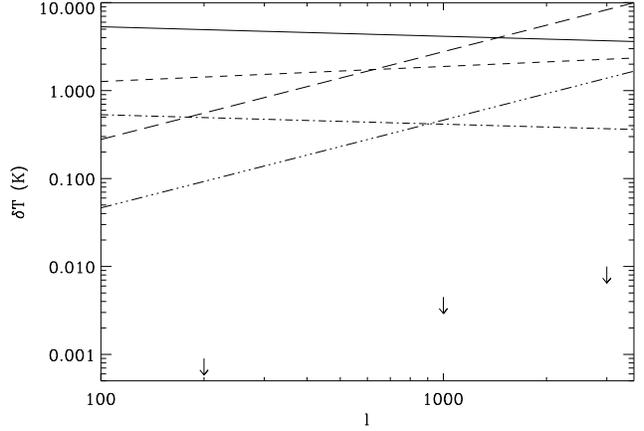}}
   \caption{Comparison between angular power spectra of various components relevant for the EoR detection. From top to bottom: total intensity (solid line) and polarization (short--dashed line) power spectra fitted from the Fan data and extrapolated down to $\sim$3~arcmin, power spectrum due to unresolved point sources (long--dashed line), extrapolation at high Galactic latitudes of the best fit power spectrum of the total intensity emission (dot--dashed line), power spectrum of the thermal noise of our observations (triple--dot--dashed line), theoretical expectations for the EoR signal (Zaldarriaga et al. 2004, downward arrows).}
              \label{eor_comp}
    \end{figure}

Therefore, if we extrapolate the measured power spectrum of diffuse total intensity emission through its best power--law fit down to 5~arcmin -- around where the EoR signal is expected to peak -- we find that its level at $\ell=2160$ is $\delta T= \sqrt{\ell (\ell+1) C^I_\ell / 2\pi} \sim 5.7$~K and the corresponding $rms$ value is $\sim$18.3~K. At moderate and high Galactic latitudes this value can drop by a factor of ten or even more if favourable sky patches are identified. Figure~\ref{eor_comp} compares the measured power spectra of foregrounds and the expected EoR signal. 

It is worth noting in this context that the detection of diffuse foregrounds is already limited by source confusion at arcmin scales in an area like the Fan region, where the diffuse emission is quite bright. This indicates that source confusion could be the most serious foreground contaminant in detecting the cosmological signal at arcmin scales. In Figure~\ref{eor_comp} we plotted the power spectrum due to residual point sources assuming that its normalization is given by the $rms$ value of $\sim$3~mJy~beam$^{-1}$ found in Section~\ref{diff_gal}. This contribution might dominate completely over the diffuse emission at angular scales smaller than $\sim$5~arcmin at all Galactic latitudes. 

Future experiments aimed at detecting the EoR signal would therefore benefit from the inclusion of long baselines in order to decrease the confusion limit.
%
%

Finally, we have already argued that the polarization signal has also to be considered in the context of the EoR detection, because a polarized signal coming from the sky will leak into Stokes $I$, inducing artificial fluctuations on top of the cosmological ones if the calibration is not perfect. Moreover, the leaked signal would vary with frequency because of Faraday rotation, generating a signal which would have characteristics similar to the cosmological one and, therefore, be extremely difficult to remove.

The seriousness of this phenomenon depends upon the strength of the polarized signal coming from the sky, its RM -- higher RM values will induce more rapid variations across the bandwidth -- and the calibration accuracy.

Recent observations at 150~MHz with the GMRT have found a region of the sky $1^\circ$ wide where the polarized brightness temperature is less than 1~K down to $\ell \sim 1000$ (Pen et al. 2008), suggesting that sky regions free of synchrotron polarization exist. We note however, that they do not impose constraints on 5--10~arcmin scales.

We have detected $rms$ fluctuations in the diffuse polarized emission of 7.2~K at 4~arcmin. On 5~arcmin scales the level of the polarization power spectrum is $\delta T \sim 3.3$~K. Most of the polarized emission appears at RM values [-7,1]~rad~m$^{-2}$. Such values do not generate rapidly rotating Stokes $Q$ and $U$ signals over a $\sim$2~MHz bandwidth but they can potentially mimic an EoR signal over a 5--7~MHz bandwidth.

Generalizing this result appears to be rather difficult. The Fan region is certainly a peculiar region in polarization. Its polarization power spectrum cannot be taken as representative of other sky areas since the region centred around the ring contributes most to it. 

It is reasonable to believe that the polarization power spectrum represents an upper limit on most of the sky at moderate and high Galactic latitudes, where lower values should be expected, in agreement with what the preliminary GMRT results suggested.

It is worth pointing out, however, that whereas the polarized brightness temperature in the Galactic halo should be intrinsically lower than in the Fan area, the interstellar medium is probably Faraday thin, because of lower values of the electron density and magnetic field. Therefore regions outside the plane could also be structured in Faraday depth and RM synthesis will be the way to detect potentially weak and structured polarized signals.

Future work will be dedicated to further characterizing foregrounds at higher Galactic latitude, with particular attention paid to their polarization features.  

\begin{acknowledgements}

We thank an anonymous referee for useful comments which improved the manuscript. GB thanks Ettore Carretti for useful discussions. The Westerbork Synthesis Radio Telescope is operated by ASTRON (Netherlands Foundation for Research in Astronomy) with support from the Netherlands Foundation for Scientific Research (NWO). LOFAR is being funded by the European Union, European Regional Development Fund, and by ``Samenwerkingsverband Noord-Nederland'', EZ/KOMPAS.

\end{acknowledgements}

\end{document}